\documentclass[aps, amsmath,amssymb,
superscriptaddress, showpacs,
reprint]{revtex4-1}

\usepackage{graphicx}
\usepackage{multirow}
\usepackage{upgreek}
\usepackage[linecolor=blue!20!white, bordercolor=blue!20!white,backgroundcolor=blue!20!white, textsize=small]{todonotes}\usepackage{units}
\usepackage{wrapfig}
\usepackage{subfigure}
\usepackage{hyperref}

\renewcommand{\d}{\mathrm{d}}

\def\c{{\mathrm c}}
\def\d{{\mathrm d}}
\def\e{{\mathrm e}}

\def\p{{\mathrm p}}

\def\R{{\mathrm R}}

\begin{document}

\title{Developments in laser-driven plasma accelerators}
\author{S. M. Hooker}
\email{simon.hooker@physics.ox.ac.uk.}
\affiliation{Department of Physics, University of Oxford, OX1 3PU, United Kingdom}
\affiliation{John Adams Institute for Accelerator Science, University of Oxford, Denys Wilkinson Building, Keble Road, Oxford OX1 3RH, United Kingdom}

\begin{abstract}
Laser-driven plasma accelerators provide acceleration gradients three orders of magnitude greater than conventional machines, offering the potential to shrink the length of accelerators by the same factor. To date, laser-acceleration of electron beams to particle energies comparable to those offered by synchrotron light sources has been demonstrated with plasma acceleration stages only a few centimetres long. This article describes the principles of operation of laser-driven plasma accelerators, and reviews their development from their proposal in 1979 to recent demonstrations. The potential applications of plasma accelerators are described and the challenges which must be overcome before they can become a practical tool are discussed.

\vspace{5mm}

This article has been published. Any citations should be for the original article: S.M. Hooker,  \textit{Nature Photonics} \textbf{7} 775 (2013). \href{http://dx.doi.org/10.1038/NPHOTON.2013.234}{DOI:10.1038/NPHOTON.2013.234}
\vspace{5mm}

\end{abstract}

\maketitle

\section{Introduction}
Particle accelerators are a ubiquitous tool in science.  The huge particle colliders --- such as the LHC at CERN --- are currently very much in the public eye, but accelerators lie at the heart of many other scientific and medical instruments. These include synchrotrons and free-electron lasers, which convert beams of energetic electrons to radiation ranging from terahertz frequencies to the x-ray region; and neutron beam facilities, which are driven by accelerated proton beams. Accelerators probe matter in all its states, enabling advances in engineering, medicine, and in the biological and physical sciences. They also have widespread applications in medicine, ranging from  x-ray diagnostics to particle beam therapies.

Virtually all of today's accelerators accelerate charged particles in the electric fields established between conducting electrodes or excited in an electromagnetic cavity; with this  approach electrical break-down limits the maximum  field to less than $\unit[100]{MV m}^{-1}$. Plasma accelerators  can  generate accelerating fields a thousand times greater, reducing  the required accelerator length by the same factor. The prospect of shrinking kilometre-scale particle colliders to a few metres, or the hundreds of metres of accelerator  within a synchrotron to a few centimetres drives this vibrant research area.

In this article we outline the physics of laser-driven plasma accelerators and discuss some of the key developments in this field. We also consider some of their potential applications and discuss the challenges which must be overcome before they can be realized. The electric fields excited in plasma by laser pulses or particle beams have been used to accelerate positive ions, protons, and electrons. We will concentrate on laser-driven acceleration of electrons but, despite this focus, the summary presented here is far from a comprehensive review. The physics of laser-driven plasma accelerators has been reviewed  in detail by Esarey et al.\ \cite{Esarey:2009}; several technical reviews of the field are available \cite{Malka:2008, Norreys:2009,Joshi:2010,Malka:2012}, as well as articles of interest to the general reader \cite{Joshi:2006, Patel:2007, Leemans:2009}.

\section{Principles of plasma accelerators}
\subsection{Key concepts}
Before describing the various regimes in which plasma accelerators operate we first summarize some concepts common to all such schemes.

\subsubsection{Overview}
\begin{figure*}[t]
\centering
\subfigure[]{\includegraphics[width = 0.45 \linewidth]{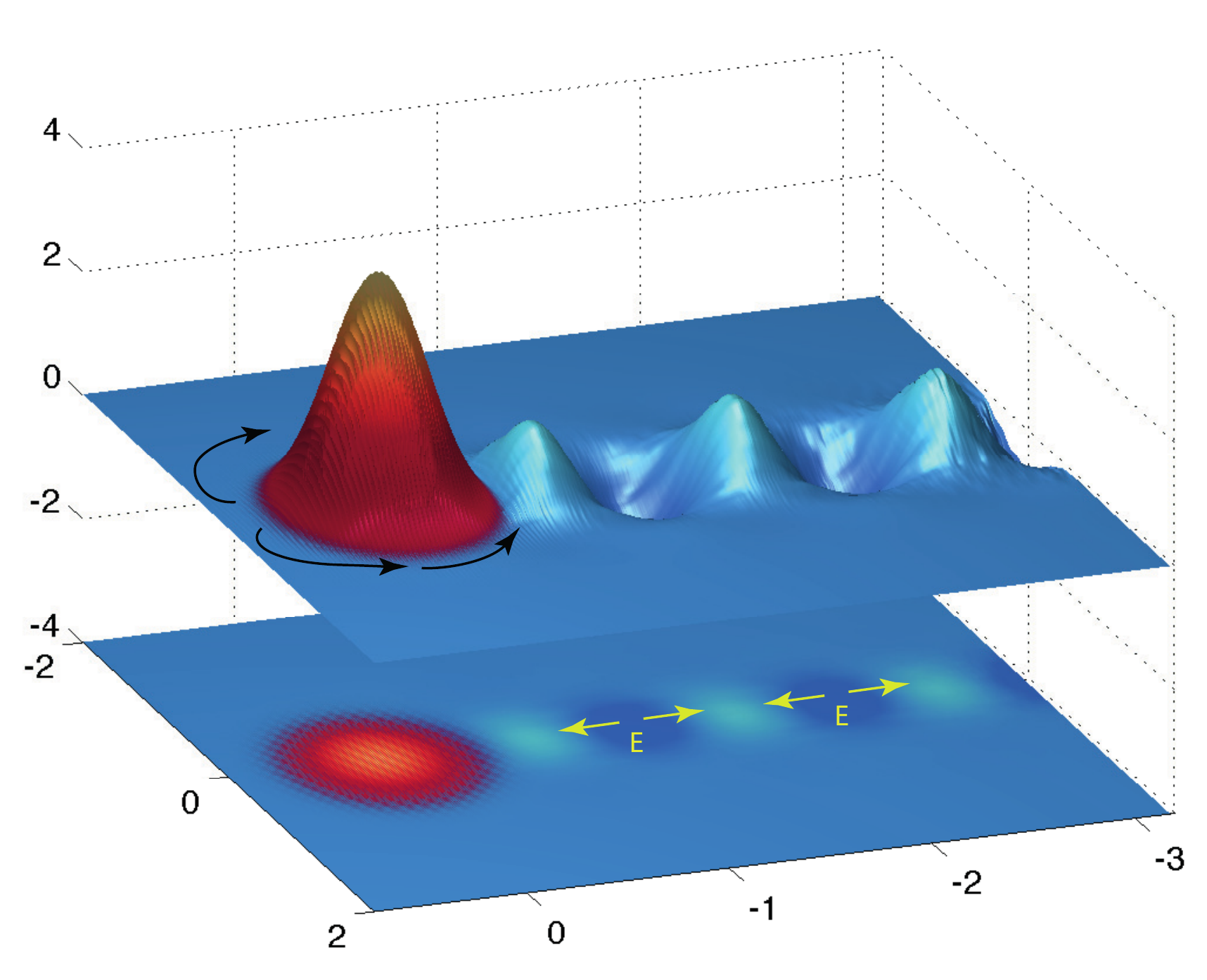}\label{Fig:Plasma_wave_linear}}
\subfigure[]{\includegraphics[width = 0.45 \linewidth]{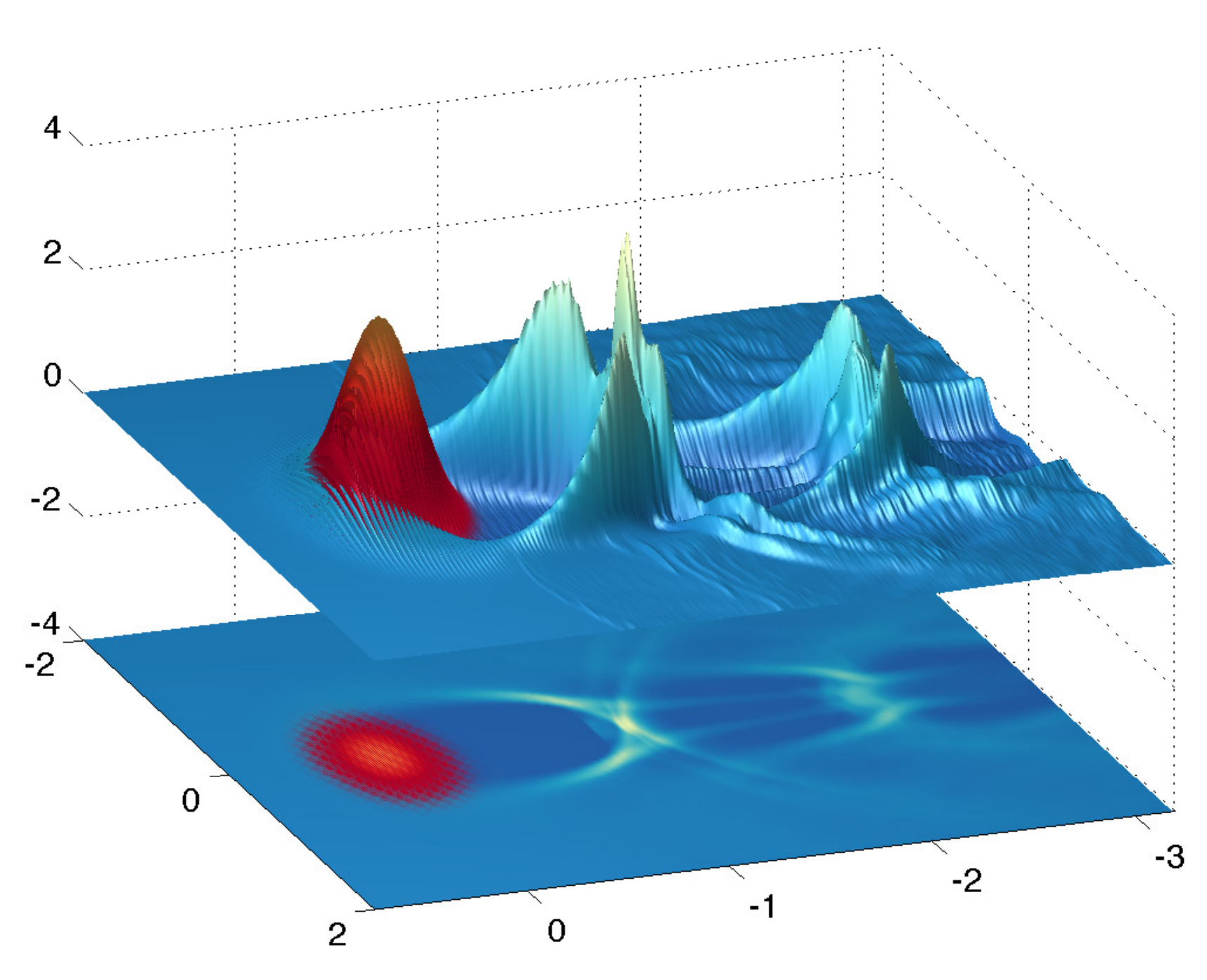}\label{Fig:Plasma_wave_nonlinear}}
\caption{Plasma waves driven by intense laser pulses. The laser pulse (red-yellow) propagates from right to left and excites a trailing plasma wave. The plasma wave amplitude (blue-green) is shown  for laser pulses with initial values of the normalized laser intensity parameter of: (a) $a_0 = 0.5$, corresponding to the linear regime; and (b) $a_0 = 4.0$, close to the bubble regime --- as indicated by the excitation of a highly nonlinear plasma wave and the formation of a `cavity' immediately behind the laser pulse. The spatial co-ordinates are in units of the plasma wavelength; the vertical scale is in arbitrary units, but that in (a) has been magnified by a factor of 10 relative to that in (b). In (a) the path of plasma electrons pushed by the ponderomotive force of the laser is indicated by the black arrows and the longitudinal electric field within the plasma wave is shown in yellow. These simulations were performed using the OSIRIS 2.0 code \cite{OSIRIS}.}
\label{Fig:Plasma_wave}
\end{figure*}

When an intense laser pulse propagates through a plasma (in simple terms, an ionized gas), the free electrons in the plasma are pushed away from the laser pulse by the ponderomotive force (see below).  The electrons are pulled back towards their original positions by the positive ions, but their momentum causes them to over-shoot and subsequently to oscillate about their initial position. The separation of electrons and ions forced by  the laser pulse therefore excites a longitudinal electron density wave --- known as a `plasma wakefield' --- which trails the driving laser pulse in a similar manner to the water wake following a moving boat. As an example, Fig.\ \ref{Fig:Plasma_wave} shows the formation of `linear' and `nonlinear' plasma wakefields for laser pulses with, respectively, relatively low and relatively high peak intensity.

The plasma oscillates at its characteristic frequency,  the plasma frequency, given (in angular units) by $\omega_\p = \sqrt{n_\e e^2 / m_\e \epsilon_0}$, where $n_\e$ is the initial electron density. The electrons are kicked forwards and backwards respectively by the front and back of the laser pulse, and hence the wakefield will be strongest when these kicks add in phase --- i.e.\ if the duration $\tau$ of the laser pulse satisfies the  condition $\omega_\p \tau \approx 1$.

The separation of electrons and ions within the plasma wave sets up huge electric fields which can be used to accelerate charged particles. Importantly,  the wakefield  propagates at the speed of the laser pulse and so it can be used to accelerate particles to relativistic energies.

The magnitude of the electric field within the plasma wave can reach the order of the `cold wave-breaking limit' $E_\mathrm{wb} = m_\e c \omega_\p /e$. For example, for a plasma density of $n_\e = 1 \times 10^{18}\unit{cm}^{-3}$, we find $\omega_\p = 5.6\times 10^{13}\, \unit{rad\, s}^{-1}$ and $E_\mathrm{wb} = \unit[96]{GeV m}^{-1}$; the plasma period $T_\p = 2 \pi/ \omega_\p \approx \unit[110]{fs}$, and hence resonant excitation requires laser pulses with a duration somewhat less than \unit[100]{fs}.

\subsubsection{The ponderomotive force}
The wakefield is driven by the ponderomotive force, which is the time-averaged force experienced by a charged particle as it moves in the fields of an electromagnetic \emph{pulse} \cite{Kruer:1988, Gibbon:2005, Esarey:2009}. One way of seeing how this force arises is to consider the mean kinetic energy of an electron oscillating in the laser field. This ponderomotive energy, $U_\p$, is proportional to the laser intensity $I$. Spatial variations in the laser intensity, and hence in $U_\p$, correspond to the ponderomotive force $F_\p = - \nabla U_\p \propto - \nabla I$. It acts to push electrons away from regions of high gradients in the laser intensity.

\subsubsection{Limits to the acceleration}
A fundamental limit to the distance over which particles can be accelerated arises from the fact that the plasma wave travels at the speed of the laser pulse --- which is less than $c$, the speed of light in vacuum  --- whereas the accelerated particles reach speeds very close to $c$. In the linear regime particles move from a region of acceleration to one of deceleration after the dephasing length  $L_\d \approx \lambda_\p^3  / \lambda^2$, where $\lambda_\p \approx 2 \pi c/\omega_\p$ and $\lambda$ are respectively the wavelengths of the plasma wave and the  driving laser. For an \unit[800]{nm} laser pulse propagating in a plasma with $n_\e = 10^{18}\unit{cm}^{-3}$, $L_\d \approx \unit[60]{mm}$.

Other limitations arise from: depletion of the driving laser energy, by transfer to the plasma wave; and defocusing of the pump laser.

\subsubsection{Trapping}
The electrons to be accelerated can either be injected into the plasma wave, or they can be pulled out of the target plasma. At low laser intensities the  wakefield has a low amplitude, and in this case the plasma electrons cannot gain sufficient momentum to become `trapped' in the plasma wave. In this regime trapping requires that the electrons are injected into the plasma wave with a sufficiently high initial energy --- just as a surfer has to swim fast enough to catch a wave. For very intense laser pulses, however, the plasma waves are highly nonlinear, and in this case some of the background plasma electrons can be trapped and accelerated; this simplifies generation of an electron beam, but at the cost of reduced control.

\subsection{Regimes of operation}
When describing the regimes in which plasma accelerators operate it is convenient to introduce  the normalized vector potential $a = e A / m_\e c$, where $A$ is the vector potential of the laser field. For a linearly-polarized pulse the  normalized vector potential at the peak of the laser pulse is  given by $a_0 \approx 0.855\sqrt{I_{18}  \lambda^2_{\unit{\mu m}}}$, where $I_{18}$ is the peak laser intensity in units of $10^{18}\unit{W cm}^{-2}$ and  $\lambda_{\unit{\mu m}}$ is the laser wavelength in  ${\unit{\mu m}}$.  This  description of the laser field is useful since the  quiver motion of the plasma electrons becomes relativistic when $a_0 \gtrsim 1$; for a laser pulse with $\lambda = \unit[1]{\mu m}$, the onset of relativistic electron motion occurs for peak intensities of approximately $1.4 \times 10^{18}\unit{W cm}^{-2}$.

\subsubsection{Linear and nonlinear regimes}
For low laser intensities ($a_0 \ll 1$) the wakefields are sinusoidal with a wavelength  $\lambda_\p = 2 \pi c/\omega_\p$, as seen in Figs \ref{Fig:Plasma_wave_linear} and \ref{Fig:Wakefield_cartoon}(a), and the  amplitude of the plasma wave and the accelerating electric field are both proportional to the laser intensity.

The velocity of the oscillating electrons increases with laser intensity. At high  intensities ($a_0 \gtrsim 1$), the relativistic increase in electron mass decreases the plasma frequency; this increases the wavelength of the plasma wave and causes the wakefield to develop a  `sawtooth' profile. These effects cause the plasma wavefronts to curve, as seen in Figure \ref{Fig:Plasma_wave_nonlinear}, as a result of the variation of the laser intensity with transverse position.

\subsubsection{Laser guiding}
One factor which determines the length over which the acceleration can be driven is the distance over which the laser intensity can be maintained. For a Gaussian laser beam of waist $w_0$ (the radius at which the beam intensity is $1/\e^2$ of its axial value), the  intensity halves at a distance equal to one Rayleigh range $z_\R = \pi w_0^2 / \lambda$ beyond the focus \cite{Hooker:2010}. Reaching the high intensities required for plasma acceleration (assuming a realistic laser power)  requires a focal spot size of order $\unit[10]{\mu m}$, corresponding to a Rayleigh range of \unit[0.3]{mm} for a $\lambda = \unit[1]{\mu m}$ laser. To drive plasma acceleration over more than a few millimetres, it is therefore necessary to guide the laser pulse.

Nature unexpectedly helps through a process known as relativistic self-focusing. Since the intensity of the laser radiation is greater on axis than in its transverse wings, the relativistic factor $\gamma$  of the plasma electrons oscillating in the laser field decreases with distance $r$ from the propagation axis. Since the refractive index of a plasma  $\eta (r)= \sqrt{1 - n_e^2 e^2 / \gamma (r) m_e \epsilon_0 \omega^2}$, this transverse variation of $\gamma$ increases the refractive index  near the axis relative to that in the wings. Just as in a gradient refractive index (GRIN) optical fibre, a refractive index profile of this form focuses the beam and so can overcome diffraction. For a steady-state beam, this relativistic self-focusing is balanced \cite{Sprangle:1987} by diffraction at the critical power  $P_\c \approx 17.4 \left(\omega / \omega_\p\right)^2 \unit{GW}$. For laser \emph{pulses} the picture is more complicated \cite{Sprangle:1990a, Sprangle:1992}, but relativistic self-focusing certainly leads to a considerable extension of the laser-plasma interaction, and has been exploited in many plasma acceletor experiments.

A plasma channel provides an alternative way of extending the distance over which the laser remains focused. Here, a  plasma column is formed in which the density is lower near the axis, creating an axially-peaked  refractive index profile and hence a guiding structure. Plasma channels suitable for guiding high-power laser pulses have been investigated by several groups, including the author's \cite{Durfee:1993, Ehrlich:1996, Hosokai:2000, Spence:2001, Gonsalves:2007, Lopes:2003}.

\section{An overview of progress}
The genesis of plasma accelerators was the pioneering 1979  paper by Tajima and Dawson \cite{Tajima:1979} in which they suggested that the strong electric fields formed within laser-driven plasma waves could accelerate charged particles to relativistic energies. Their principal idea was the excitation of relativistic plasma waves by a single, intense laser pulse, with a pulse duration somewhat less than the plasma period --- a scheme now known as the `laser wakefield accelerator' (LWFA). Here we summarize some of the key stages in the subsequent development of plasma accelerators.

\begin{figure}[tb]
\centering
\includegraphics[width = 0.75 \linewidth]{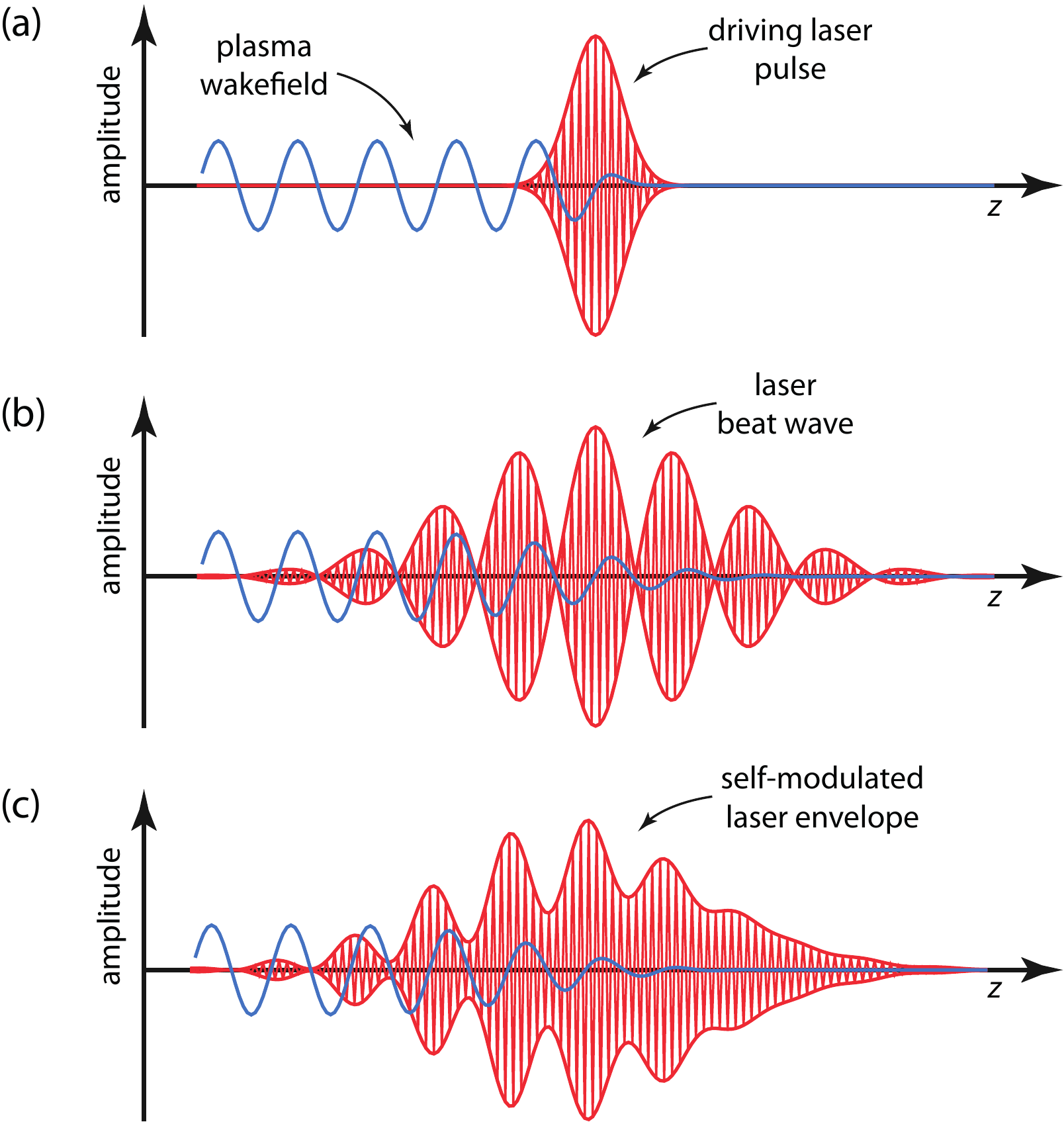}
\caption{Laser-driven plasma acceleration schemes. The amplitudes of the driving laser field and the electron density wave are shown for the cases of: (a) a single driving pulse, known as the `laser wakefield accelerator' (LWFA); (b) beating of two laser fields with a frequency difference equal to the plasma frequency, known as the `plasma beat-wave accelerator' (PBWA); (c)  self-modulation of a long laser pulse by its interaction with the plasma, known as the `self-modulated laser wakefield accelerator' (SM-LWFA). In all cases the laser pulse propagates towards positive $z$.}
\label{Fig:Wakefield_cartoon}
\end{figure}

\subsection{Plasma beat-wave accelerators}

Tajima and Dawson recognized that efficient excitation of the plasma wave would require an intense laser pulse with a duration less than  \unit[100]{fs}, and that this was demanding for the glass laser systems then available. They therefore suggested an alternative scheme: the plasma beat-wave accelerator (PBWA). This uses the fact that two co-propagating laser pulses with angular wavenumber and frequencies $k_{1,2}$ and $\omega_{1,2}$ will beat to give a modulated laser  \emph{amplitude} of the form $\cos \left[ \frac{1}{2}(\Delta k z - \Delta \omega t)\right]$, where $\Delta  \omega  = \omega_2 - \omega_1$ and $\Delta k = k_2 - k_1$. Resonant excitation of a plasma wave then occurs when the ponderomotive kicks from the modulated intensity profile of the pulse add in phase, i.e.\ when $\Delta \omega =  \omega_\p$, as shown schematically in Fig.\ \ref{Fig:Wakefield_cartoon}(b).

Beat-wave excitation of plasma waves was first demonstrated \cite{Clayton:1985} by the UCLA group in 1985. The laser used was a CO$_2$ laser \cite{Hooker:2010} delivering pulses with an energy of \unit[16]{J} and of approximately \unit[2]{ns} duration, operating simultaneously on two ro-vibrational lines with wavelengths of $\unit[9.56]{\mu m}$ and $\unit[10.59]{\mu m}$; the frequency difference of the two laser lines corresponds to a resonant  density of $n_\e = \unit[1.17 \times 10^{17}]{cm}^{-3}$.  Other groups exploited the small difference in operating wavelength of laser ions doped into different crystal hosts \cite{Amiranoff:1992}.

The first unambiguous demonstration of acceleration of electrons in a PBWA was obtained by the UCLA group in 1993, again using a dual-wavelength CO$_2$ laser \cite{Clayton:1993}. In that work \unit[70]{J}, \unit[300]{ps}  laser pulses were used to accelerate \unit[2.1]{MeV} electrons injected from a radio-freqency (RF) linac to energies up to \unit[9.1]{MeV}. Later work \cite{Tochitsky:2004} by the Brookhaven group employed a plasma channel to extend the length of the accelerator, allowing acceleration of electrons up to \unit[50]{MeV}.

The PBWA scheme suffers from an intrinsic limitation: as the amplitude of the plasma wave is increased the relativistic increase in electron mass reduces the plasma frequency; this shifts the laser beat-wave  out of resonance and leads to saturation of the plasma wave amplitude \cite{Rosenbluth:1972}. This effect, and the onset of instabilities \cite{Amiranoff:1992, Deutsch:1991, Esarey:2009},  has prevented further progress with this  scheme.

\subsection{Self-modulated LWFA}
Difficulties of this type are avoided in the self-modulated LWFA (SM-LWFA), first investigated theoretically by Andreev et al.\ \cite{Andreev:1992} and Krall et al.\ \cite{Krall:1993}. In this approach an intense laser pulse with a length $c \tau \gg \lambda_\p$ is modulated at the plasma frequency by the laser-plasma interaction, as shown in Fig.\ \ref{Fig:Wakefield_cartoon}(c). Importantly, the modulation is automatically resonant since it is driven by the local oscxillation of the xplasma; this means that resonance is maintained over the whole driving pulse even though the plasma frequency varies as a result of the changing relativistic mass shift.

\begin{figure}[tb]
\centering
\subfigure[]{\includegraphics[width = 75mm]{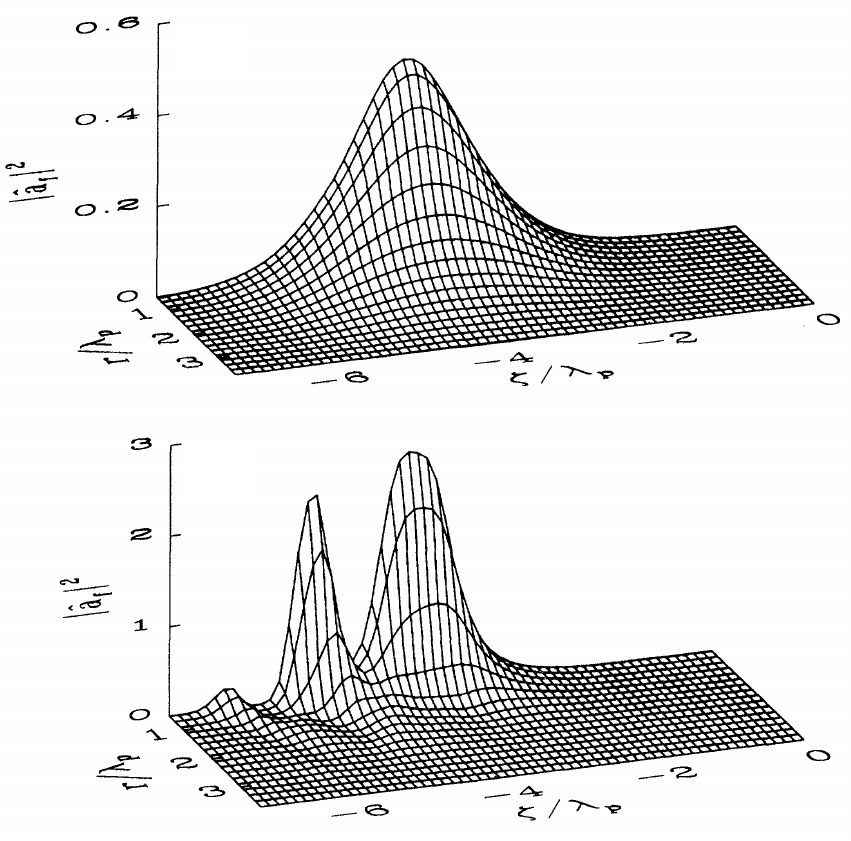}\label{Fig:Krall1993}}
\subfigure[]{\includegraphics[width = 80mm]{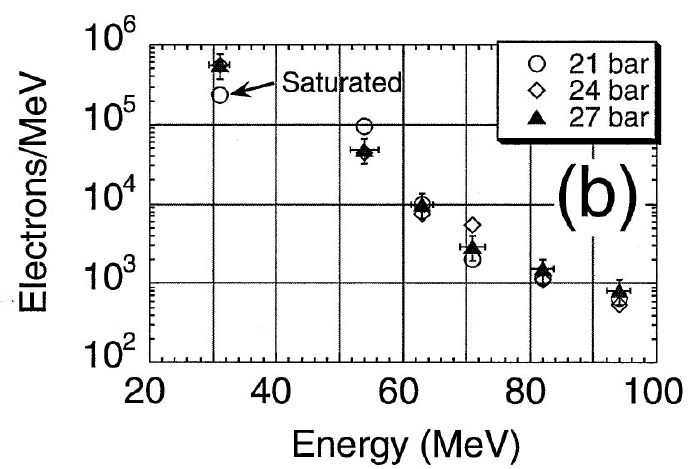}\label{Fig:Gordon1998}}
\caption{Self-modulated laser wakefield acceleration. (a) Numerical simulations of a laser pulse undergoing self-modulation as it propagates from $z = 2 z_\R$ to $z = 3.2 z_\R$, where $z_\R$ is the Rayleigh range of the input beam. In these plots the laser propagates from left to right. (After \cite{Krall:1993}. \copyright 1993 by the American Physical Society.) (b) Measured electron energy spectra measured in a forward $f/100$ cone angle at three gas jet backing pressures for \unit[20]{J}, \unit[1]{ps} laser pulses focused to a vacuum peak intensity of $\unit[6 \times 10^{18}]{W cm^{-2}}$ ($a_0 \approx 2$). The horizontal error bars indicate the range of energies incident on each detector as well as taking into account possible positioning errors. The vertical error bars reflect the uncertainty in detector sensitivity. The signal to noise ratio is independent of this error. [After \cite{Gordon:1998}. \copyright 1998 by the American Physical Society]}
\label{Fig:SMLWFA}
\end{figure}

The modulation is caused by two mechanisms: (i)  focusing (de-focusing) in regions of lower (higher) axial density in the co-propagating plasma wave excited by the leading edge of the laser pulse  \cite{Andreev:1992, Krall:1993}; (ii) beating between the driving pulse and forward Raman-scattered waves of frequencies $\omega_0 \pm \omega_\p$. The modulated pulse envelope excites a plasma wave as in the PBWA, and positive feedback increases the amplitudes of the plasma wave and the envelope modulation as they co-propagate. Figure \ref{Fig:Krall1993} shows simulations of the modulation of a long, intense laser pulse propagating through a plasma.

The first experiments on SM-LWFA were undertaken in the early 1990s, using Nd:Glass lasers \cite{Hooker:2010} delivering picosecond laser pulses with energies of 10s of joules, interacting with plasmas with densities of order $\unit[10^{19}]{cm}^{-3}$  \cite{Nakajima:1995, Modena:1995, Wagner:1997, Gordon:1998}. Figure\ \ref{Fig:Gordon1998} shows an example of an electron energy spectrum obtained in this regime by a collaboration of groups from  UCLA, Imperial College, and Ecole Polytechnique using the VULCAN laser at Rutherford Appleton Laboratory (RAL). The spectrum is seen to be broad, extending to approximately \unit[94]{MeV}.

\subsection{The bubble regime}
With the  development of high-power, short-pulse lasers based on chirped-pulse amplification (CPA)  \cite{Strickland:1985, Mourou:1998} it became realistic to consider driving a plasma wakefield  with a single, very intense laser pulse with $a_0 \gg 1$. This work built on analogous work by Rosenzweig et al.\  who considered  plasma waves driven by short, dense particle beams \cite{Rosenzweig:1991} and found that in this `blow-out' regime the plasma electrons are completely expelled from the axial region, leaving a cavity within which is formed an ideal field structure for accelerating electron bunches (see Fig.\ \ref{Fig:Plasma_wave_nonlinear}).

Following work by Mora and Antonsen \cite{Mora:1996} on the propagation of intense ($a_0 = 1$), but long ($c \tau \gg \lambda_\p$) laser pulses, Pukhov and Meyer-ter-Vehn \cite{Pukhov:2002} studied the case of wakefields driven by intense, short laser pulses. In what they termed the `bubble regime', Pukhov and Meyer-ter-Vehn  found that  some of the electrons which propagated around the edge of the bubble could be trapped at its rear, near the axis. Further, they found that the energy spectrum of the accelerated electrons could  be strongly peaked, unlike the quasi-thermal spectra generated with SM-LWFA.

\begin{figure}[tb]
\centering
\subfigure[]{\includegraphics[width = 70mm]{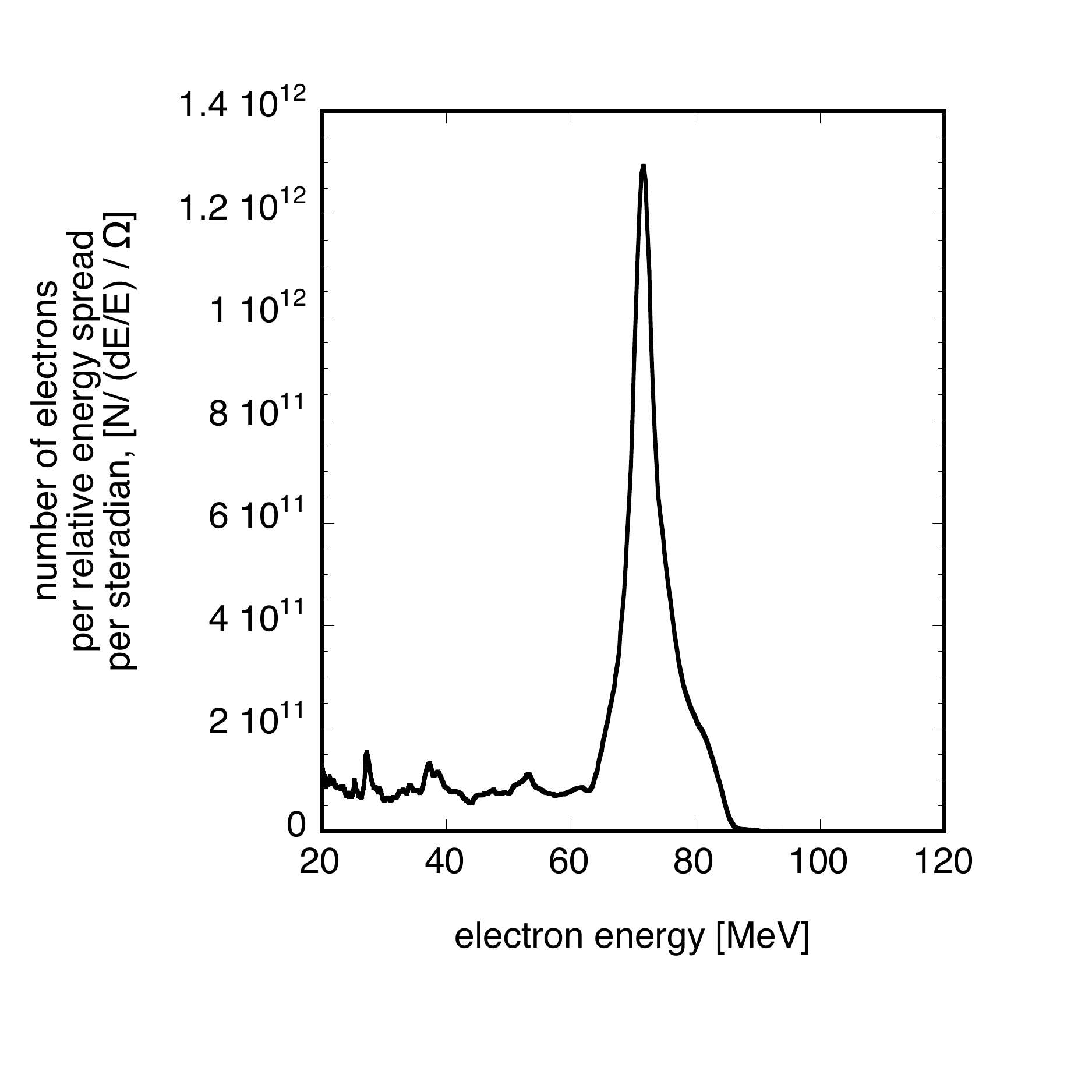}\label{Fig:Mangles2004}}
\subfigure[]{\includegraphics[width = 80mm]{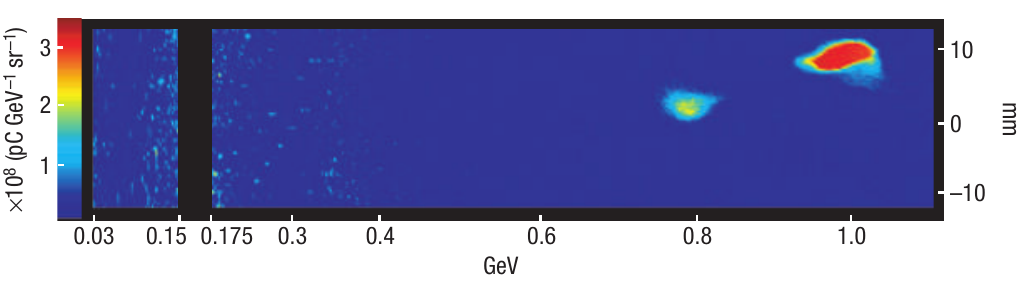}\label{Fig:Leemans2006}}
\caption{Generation of quasi-monoenergetic electron beams. (a) Measured electron energy spectrum showing acceleration to close to \unit[100]{MeV} for laser pulses of \unit[500]{mJ}, $\tau = \unit[40]{fs}$ laser pulses focused to a peak intensity of $\unit[2.5 \times 10^{18}]{W\,cm^{-2}}$ into a plasma density of $\unit[2 \times 10^{19}]{cm^{-3}}$. The energy width was $\pm 3\%$. [After \cite{Mangles:2004}.] (b) Generation of a \unit[1]{GeV} beam (and a second beam at \unit[0.8]{GeV}) by a plasma accelerator driven by a \unit[1.6]{J}, $\tau = \unit[38]{fs}$ laser pulse focused into the plasma channel formed in a capillary discharge waveguide. The axial plasma density was $\unit[4.3 \times 10^{18}]{cm^{-3}}$. The black stripe indicates an energy range not recorded by the spectrometer. [After \cite{Leemans:2006}.]}
\label{Fig:Beams}
\end{figure}

The bubble regime is of great importance since it provides a relatively straightforward way of generating  high-energy, quasi-monoenergetic electron beams without the complications of  injecting electrons from an external source or providing an external guiding structure for the driving laser pulse. The first experimental results obtained in, or close to, the bubble regime were reported in 2004 by groups at Imperial College \cite{Mangles:2004}, Lawrence Berkeley National Laboratory (LBNL) \cite{Geddes:2004}, and Laboratoire d'Optique Appliqu\'ee (LOA) \cite{Faure:2004}. In these experiments the driving lasers were  Ti:sapphire CPA systems \cite{Hooker:2010} delivering laser pulses with an energy in the range 0.5 - \unit[1]{J} and of duration 33 - \unit[55]{fs}. The targets were supersonic H$_2$ or He gas jets, ionized by the driving laser pulse to give fully-ionized plasmas of density $n_\e \approx \unit[2 \times 10^{19}]{cm}^{-3}$.  The key finding of  these experiments was the observation of electron beams with much narrower energy spectra than previously observed. Fig.\ \ref{Fig:Mangles2004} shows the result obtained by the Imperial College group using the Astra laser at RAL. Similar results were obtained by the LBNL group, who reported a spectrum peaked at \unit[86]{MeV} with a width of $< 2\%$, and by the LOA group, who measured a beam at \unit[170]{MeV} with a relative energy spread less than the spectrometer resolution of 24\%. For the first time, laser-driven plasma accelerators had generated near-monoenergetic beams, and  in so doing --- as evidenced by their appearance on the front cover of the `Dream Beam' issue of \emph{Nature}  ---  they had `arrived'.

\subsection{Plasma acceleration in waveguides}
Since the energy gained in accelerating over one dephasing length $L_\d$ varies as $1/n_e$, and $L_\d \propto n_e^{-3/2}$,   increasing the electron energy above the 100-\unit[200]{MeV} reached in the 2004 `Dream Beam' experiments requires operation at lower plasma densities and maintaining acceleration over longer distances. A distinguishing feature of the LBNL 2004 experiments is that the laser pulse was guided in the  channel \cite{Geddes:2004} formed by hydrodynamic expansion of a plasma column formed by two  laser pulses arriving before the main driving pulse. However, this technique is limited to relatively high plasma densities \cite{Durfee:1993, Geddes:2004}.

\begin{figure}[tb]
\centering
\subfigure{\includegraphics[width=80mm]{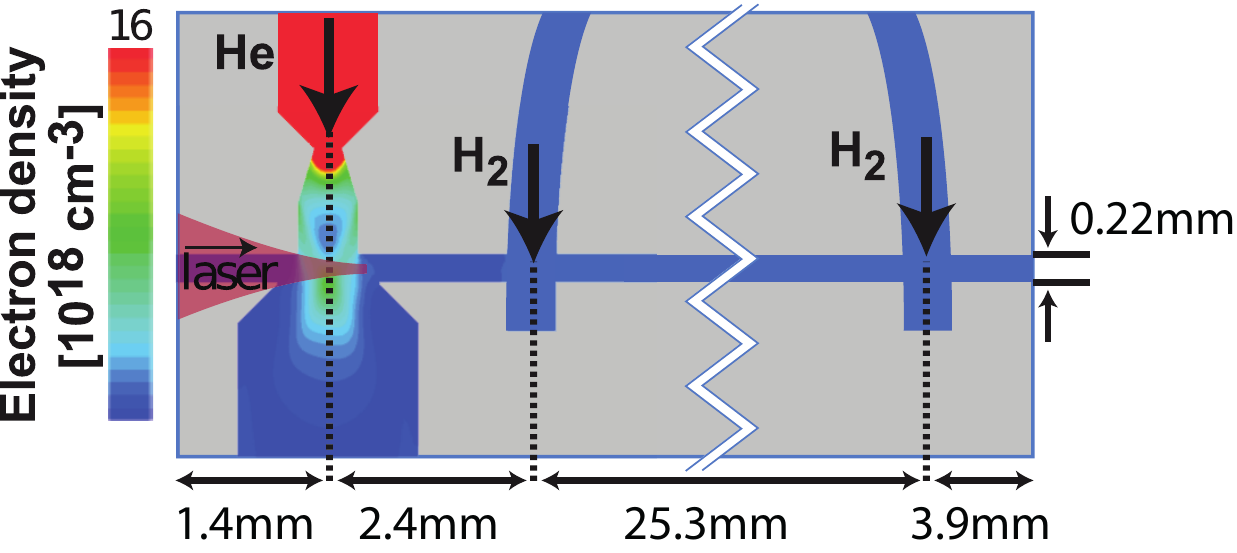}\label{Fig:Gonsalves_JetCap}}
\subfigure{\includegraphics[width=80mm]{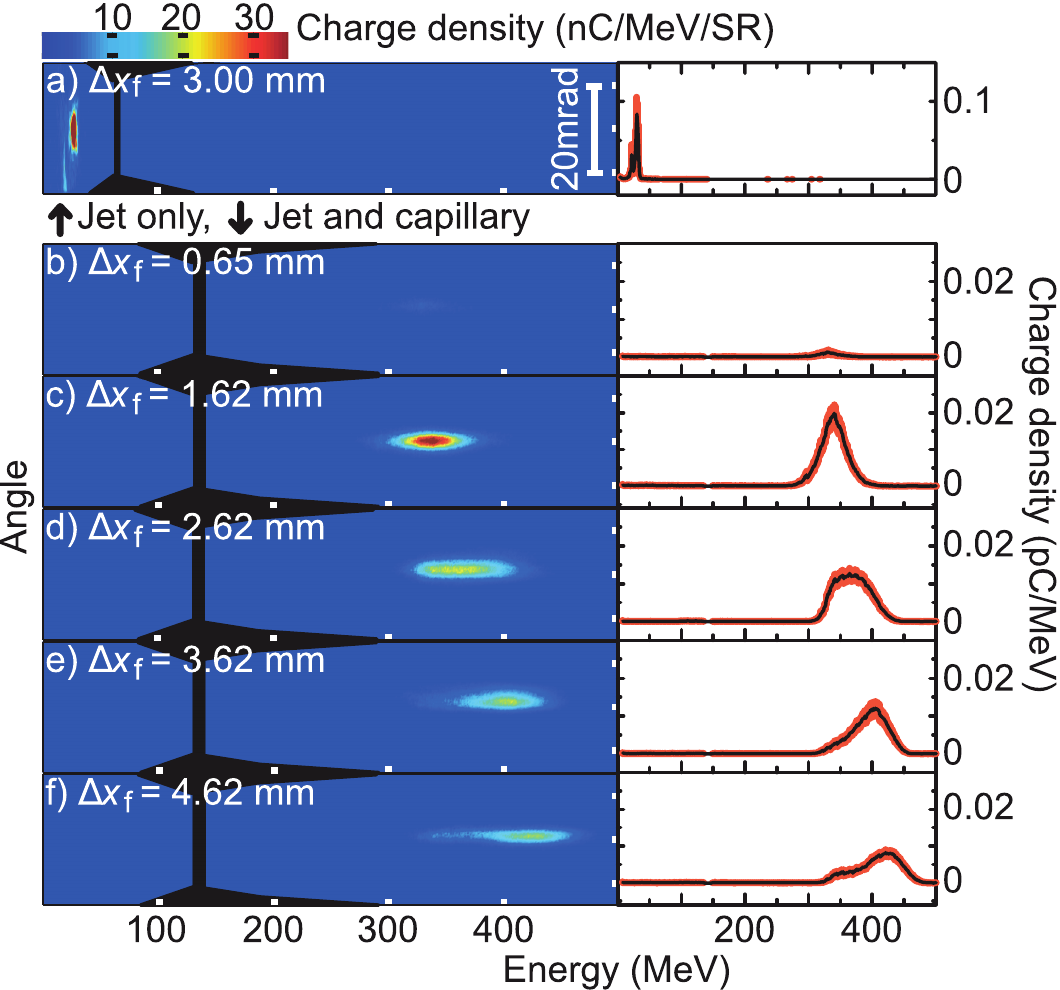}\label{Fig:Gonsalves_Spectra}}
\caption{Controlling electron injection with a density ramp. Top: schematic diagram of the combined He gas jet and hydrogen-filled capillary discharge waveguide, showing the calculated initial gas density. (a-f) Magnetic spectrometer images. The black shaded areas in each image represent the regions not covered by the spectrometer cameras. Lineouts of the mean (black curve) and the standard deviation (red area) are given to the right of each image. Note that the electron spectra are averages over 20 consecutive laser shots, demonstrating the remarkable stability of the accelerator: the RMS variations of the energy, charge and beam propagation direction were 1.9\%, 45\% (which could be improved to 6\% by changing the density profile), and \unit[0.57]{mrad} respectively. [After \cite{Gonsalves:2011}.]\label{Fig:Gonsalves2011}}
\end{figure}

A method for creating plasma channels at  lower  densities was developed by the author and his group at Oxford. In this approach, a sapphire or alumina capillary --- with a diameter of $\unit[200-300]{\mu m}$ and of length of tens of millimetres --- is filled with hydrogen gas via holes drilled near each end; Fig.\ \ref{Fig:Gonsalves_JetCap} shows a more advanced capillary design which also incorporates a gas jet for controlling electron injection (see below). The plasma is formed by pulsing a discharge through the capillary, whereupon conduction of heat to the capillary wall establishes a maximum in plasma temperature --- and hence a minimum in density ---  along the capillary axis. This allows guiding in low-Z gasses, and evolved from work by Zigler et al.\  on ablative capillary discharges where the plasma was formed from the (higher Z) wall material of the capillary \cite{Ehrlich:1996}.

 Measurements  and modelling  show that the transverse electron density profile can guide intense laser pulses with a spot size of $30-\unit[40]{\mu m}$ over tens of millimetres \cite{Spence:2001, Bobrova:2002, Butler:2002, Broks:2007,Gonsalves:2007}. In 2006, the LBNL and Oxford groups used this waveguide to guide laser pulses with an input intensity of approximately $\unit[ 3 \times 10^{18}]{W cm^{-2}}$ through a \unit[33]{mm}-long plasma channel with an axial plasma density of $\unit[4.3 \times 10^{18}]{cm^{-3}}$. As shown in Fig.\ \ref{Fig:Leemans2006}, this allowed the generation of quasi-monoenergetic beams with electron energies of \unit[1]{GeV} for the first time \cite{Leemans:2006}. Beams with electron energies up to \unit[2]{GeV} have subsequently been demonstrated by several groups employing external waveguides \cite{Karsch:2007, Ibbotson:2010} or self-guiding \cite{Kneip:2009, Pollock:2011, Mo:2012, Wang:2013}.

\section{Applications of plasma accelerators}

The potential applications of particle or radiation beams generated by laser-driven plasma accelerators include: radiography for non-destructive inspection of materials or diagnosis; radiotherapy \cite{Malka:2010}; ultrafast radiolysis studies in biology and chemistry \cite{Malka:2010, Gauduel:2010};  ultrafast studies of condensed matter;  and fundamental research, such as studies of `plasma QED'  and particle physics.

\begin{figure}[tb]
\centering
\includegraphics[width=80mm]{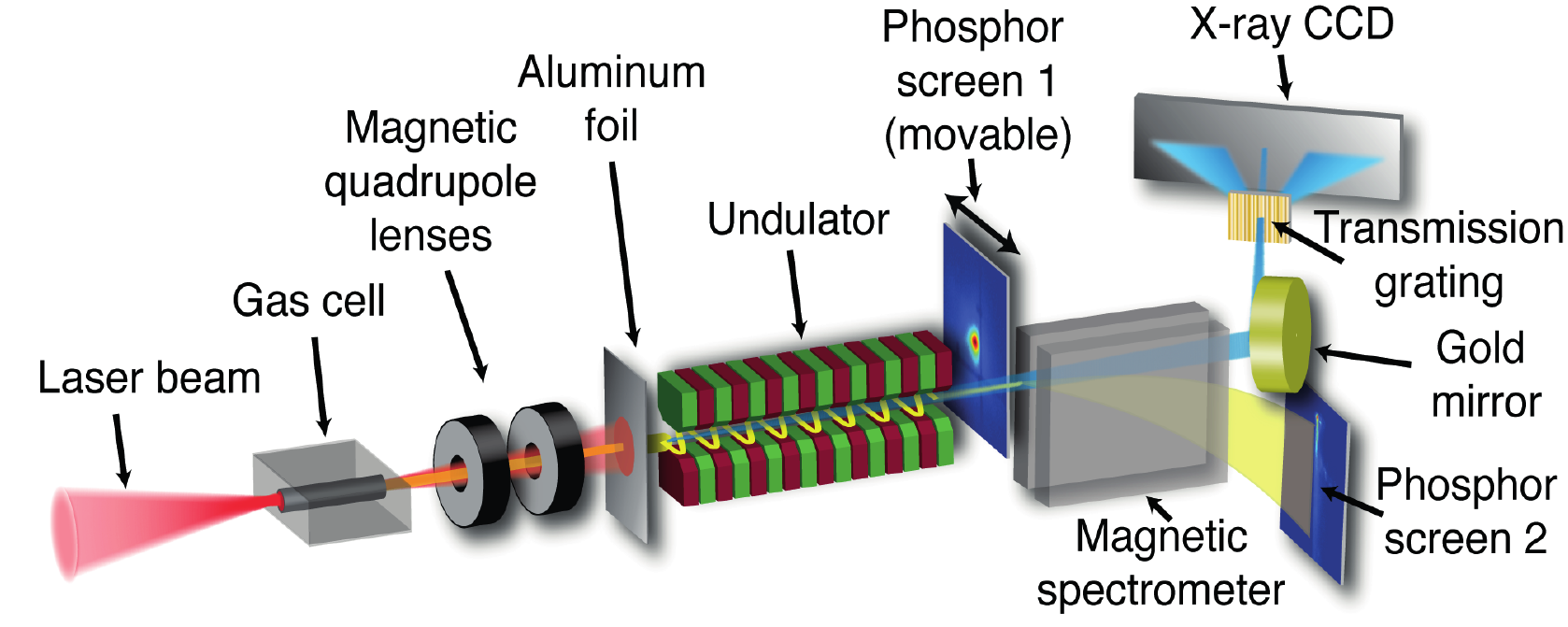}\label{Fig:Fuchs_schematic}
\caption{Generation of undulator radiation with a laser-driven plasma accelerator. A laser pulse (red) is focused into a gas cell, in which plasma waves accelerate electrons (yellow) to energies of several hundred MeV. The electron beam is collimated by a pair of quadrupole lenses. Plasma radiation and the laser beam are blocked by an aluminium foil. The electrons propagate through an undulator and emit soft-x-ray radiation into a narrow cone along the forward direction (blue). The radiation is collected by a spherical gold mirror and characterized by a transmission grating in combination with an x-ray CCD camera. Stray light is blocked by a slit in front of the grating. The pointing, divergence and spectrum of the electron beam are diagnosed by phosphor screens. This approach was used to generate soft x-rays with a wavelength of \unit[17]{nm}. [After \cite{Fuchs:2009}.]}
\label{Fig:Fuchs2009}
\end{figure}

Laser plasma accelerators are well suited to driving compact sources of radiation. Radiation has been generated at visible \cite{Schlenvoigt:2008} and soft x-ray wavelengths \cite{Fuchs:2009} by passing laser-accelerated electrons through magnetic undulators. Figure \ref{Fig:Fuchs2009} shows  results by Fuchs et al.\ \cite{Fuchs:2009} in which \unit[200]{MeV} electrons were passed through a compact (\unit[30]{cm} long) magnetic undulator to generate radiation at \unit[17]{nm} and \unit[9]{nm} (fundamental  and second-harmonic), with an estimated peak brilliance at \unit[17]{nm} of $\unit[1.3 \times 10^{17}]{PSB}$, where \unit{PSB} stands for  $\unit{photons \,s^{-1} mrad^{-2} mm^{-2} 0.1\% BW^{-1}}$.

Bright radiation is also emitted directly from the plasma accelerator as a result of the oscillation of the accelerating electron bunch in the \emph{transverse} electric field of the plasma wakefield.  These betatron oscillations generate broad-band radiation at very short wavelenths \cite{Kneip:2010, Cipiccia:2011}. For example, Kneip et al.\ \cite{Kneip:2010} have measured the betatron spectrum emitted by a \unit[250]{MeV} plasma accelerator, and showed that emission occurs in the \unit[1-100]{keV} range with a peak  brilliance of $\unit[1 \times 10^{22}]{PSB}$ at photon energies of \unit[10]{keV}. The small source size, typically a few microns,  has allowed x-ray phase contrast imaging of biological specimens \cite{Kneip:2011, Fourmaux:2011}.

In a free-electron laser (FEL), positive feedback between an electron bunch and the radiation it generates as it passes through an undulator causes exponential growth of the radiation intensity via `micro-bunching' of the electrons. FELs require very high quality electron bunches, which means that replacing the km-scale conventional accelerators of existing x-ray FELs \cite{Emma:2010} with plasma accelerators is a challenging goal. Gr\"uner et al.\ \cite{Gruner:2007} have considered the design of a FEL operating at \unit[0.25]{nm} driven by a laser-accelerated \unit[1.7]{GeV}, \unit[1.6]{nC}, \unit[10]{fs} electron bunch. Recent work has shown that the requirements on the electron bunch parameters could be eased by employing tailored undulator designs \cite{Huang:2012,Maier:2012}.

In the long  term plasma accelerators could replace the conventional accelerators in particle colliders. However, the laser and plasma target requirements for a laser-driven TeV collider are daunting. For example, Nakajima et al.\ \cite{Nakajima:2011} have considered a two-stage plasma accelerator with a plasma stage length of \unit[333]{m} driven by \unit[10.4]{kJ}, \unit[950]{fs} laser pulses with a pulse repetition rate of \unit[0.88]{kHz}; this corresponds to a \emph{mean} laser power of \unit[9.2]{MW}! An alternative design  by Schroeder et al.\ \cite{Schroeder:2012} comprises a 50-stage linac, each stage driven by \unit[32]{J}, \unit[56]{fs} laser pulses with a pulse repetition rate of \unit[15]{kHz} --- i.e.\ a mean laser power \emph{per stage} of \unit[480]{kW}.

\section{Challenges for the field}

The development of laser-driven plasma accelerators has been rapid and impressive and many groups around the world have now demonstrated electron beam generation. The parameters of these bunches can reach (although not simultaneously): particle energies up to a few GeV, relative energy spread of a few percent,  charge of \unit[10-100]{pC}, an RMS duration \cite{Tilborg:2006, Ohkubo:2007, Debus:2010, Lundh:2011} as short as  \unit[2]{fs}, and a normalized transverse emittance in the range \unit[$0.1 - 2 \pi$]{mm mrad} \cite{Fritzler:2004, Sears:2010, Brunetti:2010, Plateau:2012, Weingartner:2012, Kneip:2012}. However, a great deal remains to be done. Some of the significant challenges facing this exciting field are:

\subsection{Reduction of shot-to-shot jitter and energy spread} 
In most laser-driven plasma accelerators demonstrated to date the shot-to-shot jitter of the bunch parameters and the relative energy spread is large: relative energy spread is usually in the range 1-10\%; and typical values for shot-to-shot variation in the bunch parameters are energy 1-5\%, charge 5-50\%,  and beam pointing \unit[0.5 - 3]{mrad}. These values are very high  compared to those found in conventional accelerators. 

Improvements in the bunch parameters, and their shot-to-shot variation, could be achieved by controlling the injection and trapping of electrons into the wakefield, rather than relying on self-trapping. A wide range of ideas are being explored, including: the use of additional laser pulses \cite{Esarey:1997, Faure:2006}; slowing the plasma wakefield with a region of decreasing plasma density \cite{Bulanov:1998, Brantov:2008, Geddes:2008, Faure:2010, Gonsalves:2011};  introducing an abrupt jump in the plasma density \cite{Suk:2001, Schmid:2010}; and ionization of species doped into the plasma \cite{Rowlands-Rees:2008, Pak:2010, McGuffey:2010, Liu:2011, Pollock:2011}. 

Figure \ref{Fig:Gonsalves2011} shows one example in which  the density ramp formed at the edge of a helium gas jet embedded within a hydrogen-filled capillary discharge waveguide was used to control injection. The generated electron beams were very stable, and their energy could be adjusted between \unit[100]{MeV} and \unit[400]{MeV} by adjusting the  position of the laser focus on the density ramp.

However, although excellent progress is being made in this area, at the time of writing no  technique has emerged as the  method of choice, and none has been used to generate beams with a stability comparable to the best conventional accelerators.

\subsection{Increased particle energy} 
The next milestones for the field are the generation of \unit[10]{GeV} and \unit[100]{GeV} beams, but reaching these will require an increase in the laser pulse energy or the use of several laser pulses. The total length of the plasma accelerator must also increase: either by employing  a longer single stage; or by coupling several shorter stages \cite{Nakajima:2011, Panasenko:2010}. It may also be possible to extend the accelerator beyond the dephasing limit by employing longitudinally-profiled plasmas \cite{Sprangle:2001, Pukhov:2008}

\subsection{Increased repetition rate \& improved wall-plug efficiency} 
Many potential applications of laser-driven accelerators require  the pulse repetition rate $f_\mathrm{rep}$ to be increased.  To date, laser-driven plasma accelerators are typically operated with $f_\mathrm{rep}$ in the range \unit[0.03-1]{Hz}; this is many orders of magnitude lower than achieved, for example, by a third-generation light source.

Increasing the repetition rate from a few Hz to the kHz range will require the development of new laser technologies able to operate at substantially higher mean power and with significantly increased wall-plug efficiency. This is illustrated by the fact that \unit[100]{pC}, \unit[1]{GeV} bunches delivered at \unit[1]{kHz} have a mean \emph{particle beam} power of \unit[100]{W}. Given an efficiency for transferring energy from laser to plasma wave, and from plasma wave to particle bunch, of approximately 50\% each \cite{Schroeder:2012}, a  \emph{mean} laser power of \unit[400]{W} would be required. This is large compared to the mean power of \unit[20]{W}, or less, of present-day driving lasers --- and the requirements of a TeV  collider are even more extreme.

\subsection{Laser systems for plasma accelerators}
Progress in the development of laser-driven plasma accelerators has gone hand-in-hand with the development of laser technology, and this relationship is likely to continue. Several laser technologies have been identified  \cite{ICFA56, Dawson:2012} as being potentially important for future plasma accelerators: diode-pumped solid-state lasers, fibre lasers \cite{Richardson:2010}, and optical parametric chirped-pulse amplification (OPCPA)  \cite{Ross:1997, Dubietis:2006}. Diode-pumped continuous-wave (c.w.) lasers can provide high powers with high efficiency: coherent combination of a pair of Nd:YAG ceramic slab lasers has generated \cite{Goodno:2006} a continuous power  of \unit[19]{kW} with an optical efficiency of 30\%; and c.w.\ fibre lasers providing \unit[10]{kW} of mean power are available commercially \cite{Richardson:2010}. However, generating the ultrafast pulses required for plasma accelerators is  more challenging. Although fibre lasers have delivered pulses as short as \unit[4.5]{fs} \cite{Krauss:2009}, for pulse energies above \unit[1]{mJ} the pulse duration is typically longer than \unit[500]{fs}. For example, an ytterbium-doped, large-pitch fibre system has \cite{Eidam:2011} produced \unit[2.2]{mJ}, \unit[480]{fs} pulses at a repetition rate of \unit[5]{kHz} with an optical efficiency of 23\%. To reach the higher pulse energies needed for plasma accelerators, techniques for coherently combining the output of many fibre lasers are required; this is presently a very active area of research \cite{Morou:2013}.

An OPCPA extends conventional CPA by employing optical parametric, rather than laser, amplifiers \cite{Ross:1997, Dubietis:2006}; since no heat is deposited in the nonlinear crystal, OPCPAs can be operated at high mean powers. With OPCPA the technological challenges are largely shifted to the pump laser, although  these can be alleviated to some extent by using multiple pump beams  \cite{Tamovsauskas:2008, Alisauskas:2010, Kurita:2010}. All-OPCPA laser systems have generated laser pulses with a peak power up to \unit[16]{TW} at $f_\mathrm{rep} =  \unit[10]{Hz}$ \cite{Herrmann:2009}; a peak laser power of up to \unit[560]{TW} has been generated in a system providing one shot every \unit[30]{s} \cite{Lozhkarev:2007}, and a system with similar performance operating at $f_\mathrm{rep} =  \unit[10]{Hz}$ is planned \cite{Skrobol:2012}. OPCPA has been identified as a promising technology for `intermediate'  plasma accelerators providing GeV-scale beams at kHz pulse repetition rates \cite{ICFA56}. However, it is clear that significant further development of OPCPAs is still required.

It is worth noting that driving the wakefield with a train of laser pulses spaced by the plasma period could reduce the energy required per driving laser pulse by an order of magnitude or more. In this multiple-pulse LWFA (MP-LWFA) the spacing of the pulses can be adjusted to maintain resonant excitation, avoiding the problems of the the PBWA. This idea was explored by Nakajima et al.\  \cite{Nakajima:1992} and Umstadter et al.\ \cite{Umstadter:1994} in the early 1990s, but has only recently been re-examined \cite{Corner:2012} as a technique for driving plasma accelerators with efficient, low-energy, high repetition rate lasers.

\section{Conclusions}
Laser-driven plasma accelerators have made enormous advances since they were first proposed just over thirty years ago. Electron beams with bunch properties comparable to those used in synchrotron light sources can now be generated; albeit at lower repetition rates, with  larger energy widths and with larger shot-to-shot jitter. Several applications of laser-accelerated electron beams have already been demonstrated, and further developments in this direction can be expected.

These impressive advances have in large part been enabled by rapid progress in the development of high-power, ultra-fast laser systems and perhaps especially the development of chirped-pulse amplification. Important next chapters in the story of laser-plasma accelerators will be their successful adaptation to driving new applications in time-resolved science, and the demonstration of acceleration to particle energies of 10s and 100s of GeV. Successful delivery of these instalments will require continued  advances in high-power laser systems; close cooperation between those with expertise in lasers, plasma physics, particle physics, and conventional accelerator science; and constructive dialogue between groups working on the development of plasma accelerators and those interested in using them.

\break

\begin{acknowledgments}
I would like to thank all past and present members of my research group, and all those we have collaborated with, for stimulating discussions on the topics discussed here. I would particularly like to thank R. Bartolini, J. W. Dawson, and S. Karsch for providing material used in the preparation of in this article; N. Bourgeois for performing the simulations shown in Fig.\ \ref{Fig:Plasma_wave}, and for assistance in its preparation; and C. E.\ Webb, A. J.\ Barr and P. A.\ Walker for helpful comments on the draft article. This work was supported by the Engineering and Physical Sciences Research Council [grant no.\ EP/H011145/1] and the Leverhulme Trust [grant no.\ F/08 776/G]. I would like to acknowledge the OSIRIS Consortium, consisting of UCLA, IST (Lisbon, Portugal), and USC, for the use of OSIRIS, and IST for providing access to the OSIRIS 2.0 framework.
\end{acknowledgments}


\begin{thebibliography}{100}
\expandafter\ifx\csname url\endcsname\relax
  \def\url#1{\texttt{#1}}\fi
\expandafter\ifx\csname urlprefix\endcsname\relax\def\urlprefix{URL }\fi
\providecommand{\bibinfo}[2]{#2}
\providecommand{\eprint}[2][]{\url{#2}}

\bibitem{Esarey:2009}
\bibinfo{author}{Esarey, E}, \bibinfo{author}{Schroeder, CB} \&
  \bibinfo{author}{Leemans, WP}.
\newblock \bibinfo{title}{Physics of laser-driven plasma-based electron
  accelerators}.
\newblock \emph{\bibinfo{journal}{Rev. Mod. Phys.}}
  \textbf{\bibinfo{volume}{81}}, \bibinfo{pages}{1229--1285}
  (\bibinfo{year}{2009}).

\bibitem{Malka:2008}
\bibinfo{author}{Malka, V} \emph{et~al.}
\newblock \bibinfo{title}{{Principles and applications of compact laser-plasma
  accelerators}}.
\newblock \emph{\bibinfo{journal}{Nat. Phys.}} \textbf{\bibinfo{volume}{4}},
  \bibinfo{pages}{447--453} (\bibinfo{year}{2008}).

\bibitem{Norreys:2009}
\bibinfo{author}{Norreys, PA}.
\newblock \bibinfo{title}{{Laser-driven particle acceleration}}.
\newblock \emph{\bibinfo{journal}{Nat. Photonics}}
  \textbf{\bibinfo{volume}{3}}, \bibinfo{pages}{423--425}
  (\bibinfo{year}{2009}).

\bibitem{Joshi:2010}
\bibinfo{author}{Joshi, C} \& \bibinfo{author}{Malka, V}.
\newblock \bibinfo{title}{{Focus on Laser- and Beam-Driven Plasma
  Accelerators}}.
\newblock \emph{\bibinfo{journal}{New Journal Of Physics}}
  \textbf{\bibinfo{volume}{12}}, \bibinfo{pages}{045003}
  (\bibinfo{year}{2010}).

\bibitem{Malka:2012}
\bibinfo{author}{Malka, V}.
\newblock \bibinfo{title}{{Laser plasma accelerators}}.
\newblock \emph{\bibinfo{journal}{Phys. Plasmas}}
  \textbf{\bibinfo{volume}{19}}, \bibinfo{pages}{055501}
  (\bibinfo{year}{2012}).

\bibitem{Joshi:2006}
\bibinfo{author}{Joshi, C}.
\newblock \bibinfo{title}{{Plasma accelerators}}.
\newblock \emph{\bibinfo{journal}{Sci. Am.}} \textbf{\bibinfo{volume}{294}},
  \bibinfo{pages}{40--47} (\bibinfo{year}{2006}).

\bibitem{Patel:2007}
\bibinfo{author}{Patel, N}.
\newblock \bibinfo{title}{{Accelerator physics: The plasma revolution}}.
\newblock \emph{\bibinfo{journal}{Nature}} \textbf{\bibinfo{volume}{449}},
  \bibinfo{pages}{133--135} (\bibinfo{year}{2007}).

\bibitem{Leemans:2009}
\bibinfo{author}{Leemans, WP} \& \bibinfo{author}{Esarey, E}.
\newblock \bibinfo{title}{{Laser-driven plasma-wave electron accelerators}}.
\newblock \emph{\bibinfo{journal}{Phys. Today}} \textbf{\bibinfo{volume}{62}},
  \bibinfo{pages}{44} (\bibinfo{year}{2009}).

\bibitem{OSIRIS}
\bibinfo{author}{Fonseca, R} \emph{et~al.}
\newblock \bibinfo{title}{Osiris: a three-dimensional, fully relativistic
  particle in cell code for modeling plasma based accelerators}.
\newblock \emph{\bibinfo{journal}{Computational Science—ICCS 2002}}
  \bibinfo{pages}{342--351} (\bibinfo{year}{2002}).

\bibitem{Kruer:1988}
\bibinfo{author}{Kruer, W}.
\newblock \emph{\bibinfo{title}{The physics of laser plasma interactions}}
  (\bibinfo{publisher}{Addison-Wesley}, \bibinfo{year}{1988}).

\bibitem{Gibbon:2005}
\bibinfo{author}{Gibbon, P}.
\newblock \emph{\bibinfo{title}{Short pulse laser interactions with matter : an
  introduction}} (\bibinfo{publisher}{Imperial College Press},
  \bibinfo{address}{London}, \bibinfo{year}{2005}).

\bibitem{Hooker:2010}
\bibinfo{author}{Hooker, SM} \& \bibinfo{author}{Webb, CE}.
\newblock \emph{\bibinfo{title}{Laser Physics}} (\bibinfo{publisher}{Oxford
  University Press}, \bibinfo{address}{Great Clarendon Street, Oxford, United
  Kingdom}, \bibinfo{year}{2010}).

\bibitem{Sprangle:1987}
\bibinfo{author}{Sprangle, P}, \bibinfo{author}{Tang, CM} \&
  \bibinfo{author}{Esarey, E}.
\newblock \bibinfo{title}{Relativistic self-focusing of short-pulse radiation
  beams in plasmas}.
\newblock \emph{\bibinfo{journal}{IEEE Trans. Plasma Sci.}}
  \textbf{\bibinfo{volume}{15}}, \bibinfo{pages}{145 --153}
  (\bibinfo{year}{1987}).

\bibitem{Sprangle:1990a}
\bibinfo{author}{Sprangle, P}, \bibinfo{author}{Esarey, E} \&
  \bibinfo{author}{Ting, A}.
\newblock \bibinfo{title}{Nonlinear theory of intense laser-plasma
  interactions}.
\newblock \emph{\bibinfo{journal}{Phys. Rev. Lett.}}
  \textbf{\bibinfo{volume}{64}}, \bibinfo{pages}{2011--2014}
  (\bibinfo{year}{1990}).

\bibitem{Sprangle:1992}
\bibinfo{author}{Sprangle, P}, \bibinfo{author}{Esarey, E},
  \bibinfo{author}{Krall, J} \& \bibinfo{author}{Joyce, G}.
\newblock \bibinfo{title}{Propagation and guiding of intense laser pulses in
  plasmas}.
\newblock \emph{\bibinfo{journal}{Phys. Rev. Lett.}}
  \textbf{\bibinfo{volume}{69}}, \bibinfo{pages}{2200--2203}
  (\bibinfo{year}{1992}).

\bibitem{Durfee:1993}
\bibinfo{author}{Durfee, CG} \& \bibinfo{author}{Milchberg, HM}.
\newblock \bibinfo{title}{Light pipe for high intensity laser pulses}.
\newblock \emph{\bibinfo{journal}{Phys. Rev. Lett.}}
  \textbf{\bibinfo{volume}{71}}, \bibinfo{pages}{2409--2412}
  (\bibinfo{year}{1993}).

\bibitem{Ehrlich:1996}
\bibinfo{author}{Ehrlich, Y} \emph{et~al.}
\newblock \bibinfo{title}{Guiding of high intensity laser pulses in straight
  and curved plasma channel experiments}.
\newblock \emph{\bibinfo{journal}{Phys. Rev. Lett.}}
  \textbf{\bibinfo{volume}{77}}, \bibinfo{pages}{4186--4189}
  (\bibinfo{year}{1996}).

\bibitem{Hosokai:2000}
\bibinfo{author}{Hosokai, T} \emph{et~al.}
\newblock \bibinfo{title}{{Optical guidance of terrawatt laser pulses by the
  implosion phase of a fast Z-pinch discharge in a gas-filled capillary}}.
\newblock \emph{\bibinfo{journal}{Opt. Lett.}} \textbf{\bibinfo{volume}{25}},
  \bibinfo{pages}{10--12} (\bibinfo{year}{2000}).

\bibitem{Spence:2001}
\bibinfo{author}{Spence, DJ} \& \bibinfo{author}{Hooker, SM}.
\newblock \bibinfo{title}{Investigation of a hydrogen plasma waveguide}.
\newblock \emph{\bibinfo{journal}{Phys. Rev. E}} \textbf{\bibinfo{volume}{63}},
  \bibinfo{pages}{015401} (\bibinfo{year}{2001}).

\bibitem{Gonsalves:2007}
\bibinfo{author}{Gonsalves, AJ}, \bibinfo{author}{Rowlands-Rees, TP},
  \bibinfo{author}{Broks, BHP}, \bibinfo{author}{van~der Mullen, JJAM} \&
  \bibinfo{author}{Hooker, SM}.
\newblock \bibinfo{title}{Transverse interferometry of a hydrogen-filled
  capillary discharge waveguide}.
\newblock \emph{\bibinfo{journal}{Phys. Rev. Lett.}}
  \textbf{\bibinfo{volume}{98}}, \bibinfo{pages}{025002}
  (\bibinfo{year}{2007}).

\bibitem{Lopes:2003}
\bibinfo{author}{Lopes, NC} \emph{et~al.}
\newblock \bibinfo{title}{Plasma channels produced by a laser-triggered
  high-voltage discharge}.
\newblock \emph{\bibinfo{journal}{Phys. Rev. E}} \textbf{\bibinfo{volume}{68}},
  \bibinfo{pages}{035402} (\bibinfo{year}{2003}).

\bibitem{Tajima:1979}
\bibinfo{author}{Tajima, T} \& \bibinfo{author}{Dawson, JM}.
\newblock \bibinfo{title}{Laser electron-accelerator}.
\newblock \emph{\bibinfo{journal}{Phys. Rev. Lett.}}
  \textbf{\bibinfo{volume}{43}}, \bibinfo{pages}{267--270}
  (\bibinfo{year}{1979}).

\bibitem{Clayton:1985}
\bibinfo{author}{Clayton, CE}, \bibinfo{author}{Joshi, C},
  \bibinfo{author}{Darrow, C, C.~B} \& \bibinfo{author}{Umstadter, D}.
\newblock \bibinfo{title}{{Relativistic Plasma-Wave Excitation by Collinear
  Optical Mixing}}.
\newblock \emph{\bibinfo{journal}{Phys. Rev. Lett.}}
  \textbf{\bibinfo{volume}{54}}, \bibinfo{pages}{2343--2346}
  (\bibinfo{year}{1985}).

\bibitem{Amiranoff:1992}
\bibinfo{author}{Amiranoff, F} \emph{et~al.}
\newblock \bibinfo{title}{Observation of modulational instability in nd-laser
  beat-wave experiments}.
\newblock \emph{\bibinfo{journal}{Phys. Rev. Lett.}}
  \textbf{\bibinfo{volume}{68}}, \bibinfo{pages}{3710--3713}
  (\bibinfo{year}{1992}).

\bibitem{Clayton:1993}
\bibinfo{author}{Clayton, CE} \emph{et~al.}
\newblock \bibinfo{title}{Ultrahigh-gradient acceleration of injected electrons
  by laser-excited relativistic electron plasma waves}.
\newblock \emph{\bibinfo{journal}{Phys. Rev. Lett.}}
  \textbf{\bibinfo{volume}{70}}, \bibinfo{pages}{37--40}
  (\bibinfo{year}{1993}).

\bibitem{Tochitsky:2004}
\bibinfo{author}{Tochitsky, SY} \emph{et~al.}
\newblock \bibinfo{title}{Enhanced acceleration of injected electrons in a
  laser-beat-wave-induced plasma channel}.
\newblock \emph{\bibinfo{journal}{Phys. Rev. Lett.}}
  \textbf{\bibinfo{volume}{92}}, \bibinfo{pages}{095004}
  (\bibinfo{year}{2004}).

\bibitem{Rosenbluth:1972}
\bibinfo{author}{Rosenbluth, MN} \& \bibinfo{author}{Liu, CS}.
\newblock \bibinfo{title}{Excitation of plasma waves by two laser beams}.
\newblock \emph{\bibinfo{journal}{Phys. Rev. Lett.}}
  \textbf{\bibinfo{volume}{29}}, \bibinfo{pages}{701--705}
  (\bibinfo{year}{1972}).

\bibitem{Deutsch:1991}
\bibinfo{author}{Deutsch, M}, \bibinfo{author}{Meerson, B} \&
  \bibinfo{author}{Golub, JE}.
\newblock \bibinfo{title}{Strong plasma wave excitation by a ``chirped'' laser
  beat wave}.
\newblock \emph{\bibinfo{journal}{Phys. Fluids B}}
  \textbf{\bibinfo{volume}{3}}, \bibinfo{pages}{1773--1780}
  (\bibinfo{year}{1991}).

\bibitem{Andreev:1992}
\bibinfo{author}{Andreev, NE}, \bibinfo{author}{Gorbunov, LM},
  \bibinfo{author}{Kirsanove, VI}, \bibinfo{author}{Pogosova, AA} \&
  \bibinfo{author}{Ramazashvili, RR}.
\newblock \bibinfo{title}{{Resonant Excitation of Wakefields by a Laser-Pulse
  in a Plasma}}.
\newblock \emph{\bibinfo{journal}{{JETP} Lett.}} \textbf{\bibinfo{volume}{55}},
  \bibinfo{pages}{571--576} (\bibinfo{year}{1992}).

\bibitem{Krall:1993}
\bibinfo{author}{Krall, J}, \bibinfo{author}{Ting, A}, \bibinfo{author}{Esarey,
  E} \& \bibinfo{author}{Sprangle, P}.
\newblock \bibinfo{title}{Enhanced acceleration in a self-modulated-laser
  wake-field accelerator}.
\newblock \emph{\bibinfo{journal}{Phys. Rev. E}} \textbf{\bibinfo{volume}{48}},
  \bibinfo{pages}{2157--2161} (\bibinfo{year}{1993}).

\bibitem{Nakajima:1995}
\bibinfo{author}{Nakajima, K} \emph{et~al.}
\newblock \bibinfo{title}{Observation of ultrahigh gradient electron
  acceleration by a self-modulated intense short laser pulse}.
\newblock \emph{\bibinfo{journal}{Phys. Rev. Lett.}}
  \textbf{\bibinfo{volume}{74}}, \bibinfo{pages}{4428--4431}
  (\bibinfo{year}{1995}).

\bibitem{Modena:1995}
\bibinfo{author}{Modena, A} \emph{et~al.}
\newblock \bibinfo{title}{{Electron Acceleration From the Breaking of
  Relativistic Plasma-Waves}}.
\newblock \emph{\bibinfo{journal}{Nature}} \textbf{\bibinfo{volume}{377}},
  \bibinfo{pages}{606--608} (\bibinfo{year}{1995}).

\bibitem{Wagner:1997}
\bibinfo{author}{Wagner, R}, \bibinfo{author}{Chen, SY},
  \bibinfo{author}{Maksimchuk, A} \& \bibinfo{author}{Umstadter, D}.
\newblock \bibinfo{title}{Electron acceleration by a laser wakefield in a
  relativistically self-guided channel}.
\newblock \emph{\bibinfo{journal}{Phys. Rev. Lett.}}
  \textbf{\bibinfo{volume}{78}}, \bibinfo{pages}{3125--3128}
  (\bibinfo{year}{1997}).

\bibitem{Gordon:1998}
\bibinfo{author}{Gordon, D} \emph{et~al.}
\newblock \bibinfo{title}{Observation of electron energies beyond the linear
  dephasing limit from a laser-excited relativistic plasma wave}.
\newblock \emph{\bibinfo{journal}{Phys. Rev. Lett.}}
  \textbf{\bibinfo{volume}{80}}, \bibinfo{pages}{2133--2136}
  (\bibinfo{year}{1998}).

\bibitem{Strickland:1985}
\bibinfo{author}{Strickland, D} \& \bibinfo{author}{Mourou, GA}.
\newblock \bibinfo{title}{Compression of amplified chirped optical pulses}.
\newblock \emph{\bibinfo{journal}{Opt. Commun.}} \textbf{\bibinfo{volume}{55}},
  \bibinfo{pages}{447 -- 449} (\bibinfo{year}{1985}).

\bibitem{Mourou:1998}
\bibinfo{author}{Mourou, GA}, \bibinfo{author}{Barty, CPJ} \&
  \bibinfo{author}{Perry, MD}.
\newblock \bibinfo{title}{{Ultrahigh-Intensity Lasers: Physics of the Extreme
  on a Tabletop}}.
\newblock \emph{\bibinfo{journal}{Phys. Today}} \textbf{\bibinfo{volume}{51}},
  \bibinfo{pages}{22} (\bibinfo{year}{1998}).

\bibitem{Rosenzweig:1991}
\bibinfo{author}{Rosenzweig, JB}, \bibinfo{author}{Breizman, B},
  \bibinfo{author}{Katsouleas, T} \& \bibinfo{author}{Su, JJ}.
\newblock \bibinfo{title}{Acceleration and focusing of electrons in
  two-dimensional nonlinear plasma wake fields}.
\newblock \emph{\bibinfo{journal}{Phys. Rev. A}} \textbf{\bibinfo{volume}{44}},
  \bibinfo{pages}{R6189--R6192} (\bibinfo{year}{1991}).

\bibitem{Mora:1996}
\bibinfo{author}{Mora, P} \& \bibinfo{author}{Antonsen, TM}.
\newblock \bibinfo{title}{Electron cavitation and acceleration in the wake of
  an ultraintense, self-focused laser pulse}.
\newblock \emph{\bibinfo{journal}{Phys. Rev. E}} \textbf{\bibinfo{volume}{53}},
  \bibinfo{pages}{R2068--R2071} (\bibinfo{year}{1996}).

\bibitem{Pukhov:2002}
\bibinfo{author}{Pukhov, A} \& \bibinfo{author}{Meyer-ter Vehn, J}.
\newblock \bibinfo{title}{Laser wake field acceleration: the highly non-linear
  broken-wave regime}.
\newblock \emph{\bibinfo{journal}{Appl. Phys. B: Lasers Opt.}}
  \textbf{\bibinfo{volume}{74}}, \bibinfo{pages}{355--361}
  (\bibinfo{year}{2002}).

\bibitem{Mangles:2004}
\bibinfo{author}{Mangles, SPD} \emph{et~al.}
\newblock \bibinfo{title}{Monoenergetic beams of relativistic electrons from
  intense laser-plasma interactions}.
\newblock \emph{\bibinfo{journal}{Nature}} \textbf{\bibinfo{volume}{431}},
  \bibinfo{pages}{535--538} (\bibinfo{year}{2004}).

\bibitem{Geddes:2004}
\bibinfo{author}{Geddes, CGR} \emph{et~al.}
\newblock \bibinfo{title}{High-quality electron beams from a laser wakefield
  accelerator using plasma-channel guiding}.
\newblock \emph{\bibinfo{journal}{Nature}} \textbf{\bibinfo{volume}{431}},
  \bibinfo{pages}{538--541} (\bibinfo{year}{2004}).

\bibitem{Faure:2004}
\bibinfo{author}{Faure, J} \emph{et~al.}
\newblock \bibinfo{title}{A laser-plasma accelerator producing monoenergetic
  electron beams}.
\newblock \emph{\bibinfo{journal}{Nature}} \textbf{\bibinfo{volume}{431}},
  \bibinfo{pages}{541--544} (\bibinfo{year}{2004}).

\bibitem{Leemans:2006}
\bibinfo{author}{Leemans, WP} \emph{et~al.}
\newblock \bibinfo{title}{{GeV} electron beams from a centimetre-scale
  accelerator}.
\newblock \emph{\bibinfo{journal}{Nat. Phys.}} \textbf{\bibinfo{volume}{2}},
  \bibinfo{pages}{696--699} (\bibinfo{year}{2006}).

\bibitem{Gonsalves:2011}
\bibinfo{author}{Gonsalves, AJ} \emph{et~al.}
\newblock \bibinfo{title}{{Tunable laser plasma accelerator based on
  longitudinal density tailoring}}.
\newblock \emph{\bibinfo{journal}{Nat. Phys.}} \textbf{\bibinfo{volume}{7}},
  \bibinfo{pages}{862--866} (\bibinfo{year}{2011}).

\bibitem{Bobrova:2002}
\bibinfo{author}{Bobrova, NA} \emph{et~al.}
\newblock \bibinfo{title}{Simulations of a hydrogen-filled capillary discharge
  waveguide}.
\newblock \emph{\bibinfo{journal}{Phys. Rev. E}} \textbf{\bibinfo{volume}{65}},
  \bibinfo{pages}{016407} (\bibinfo{year}{2002}).

\bibitem{Butler:2002}
\bibinfo{author}{Butler, A}, \bibinfo{author}{Spence, DJ} \&
  \bibinfo{author}{Hooker, SM}.
\newblock \bibinfo{title}{Guiding of high-intensity laser pulses with a
  hydrogen-filled capillary discharge waveguide}.
\newblock \emph{\bibinfo{journal}{Phys. Rev. Lett.}}
  \textbf{\bibinfo{volume}{89}}, \bibinfo{pages}{185003}
  (\bibinfo{year}{2002}).

\bibitem{Broks:2007}
\bibinfo{author}{Broks, BHP} \emph{et~al.}
\newblock \bibinfo{title}{{Modeling of a square pulsed capillary discharge
  waveguide for interferometry measurements}}.
\newblock \emph{\bibinfo{journal}{Phys. Plasmas}}
  \textbf{\bibinfo{volume}{14}}, \bibinfo{pages}{023501}
  (\bibinfo{year}{2007}).

\bibitem{Karsch:2007}
\bibinfo{author}{Karsch, S} \emph{et~al.}
\newblock \bibinfo{title}{{GeV-scale} electron acceleration in a gas-filled
  capillary discharge waveguide}.
\newblock \emph{\bibinfo{journal}{New J. Phys.}} \textbf{\bibinfo{volume}{9}},
  \bibinfo{pages}{415} (\bibinfo{year}{2007}).

\bibitem{Ibbotson:2010}
\bibinfo{author}{Ibbotson, TPA} \emph{et~al.}
\newblock \bibinfo{title}{Laser-wakefield acceleration of electron beams in a
  low density plasma channel}.
\newblock \emph{\bibinfo{journal}{Phys. Rev. ST Accel. Beams}}
  \textbf{\bibinfo{volume}{13}}, \bibinfo{pages}{031301}
  (\bibinfo{year}{2010}).

\bibitem{Kneip:2009}
\bibinfo{author}{Kneip, S} \emph{et~al.}
\newblock \bibinfo{title}{{Near-GeV Acceleration of Electrons by a Nonlinear
  Plasma Wave Driven by a Self-Guided Laser Pulse}}.
\newblock \emph{\bibinfo{journal}{Phys. Rev. Lett.}}
  \textbf{\bibinfo{volume}{103}}, \bibinfo{pages}{035002}
  (\bibinfo{year}{2009}).

\bibitem{Pollock:2011}
\bibinfo{author}{Pollock, B} \emph{et~al.}
\newblock \bibinfo{title}{Demonstration of a narrow energy spread, {~0.5\,GeV}
  electron beam from a two-stage laser wakefield accelerator}.
\newblock \emph{\bibinfo{journal}{Phys. Rev. Lett.}}
  \textbf{\bibinfo{volume}{107}} (\bibinfo{year}{2011}).

\bibitem{Mo:2012}
\bibinfo{author}{Mo, MZ} \emph{et~al.}
\newblock \bibinfo{title}{{Quasimonoenergetic electron beams from laser
  wakefield acceleration in pure nitrogen}}.
\newblock \emph{\bibinfo{journal}{Appl. Phys. Lett.}}
  \textbf{\bibinfo{volume}{100}}, \bibinfo{pages}{074101}
  (\bibinfo{year}{2012}).

\bibitem{Wang:2013}
\bibinfo{author}{Wang, X} \emph{et~al.}
\newblock \bibinfo{title}{{Quasi-monoenergetic laser-plasma acceleration of
  electrons to 2\,GeV}}.
\newblock \emph{\bibinfo{journal}{Nat. Commun.}} \textbf{\bibinfo{volume}{4}},
  \bibinfo{pages}{1--9} (\bibinfo{year}{2013}).

\bibitem{Malka:2010}
\bibinfo{author}{Malka, V}, \bibinfo{author}{Faure, J} \&
  \bibinfo{author}{Gauduel, YA}.
\newblock \bibinfo{title}{{Ultra-short electron beams based spatio-temporal
  radiation biology and radiotherapy}}.
\newblock In \emph{\bibinfo{booktitle}{Mutat. Res., Rev. Mutat. Res.}},
  \bibinfo{pages}{142--151} (\bibinfo{organization}{Ecole Polytech, Ecole Natl
  Super Tech Avancees, Lab Opt Appliquee, CNRS,UMR 7639, F-91761 Palaiseau,
  France}, \bibinfo{year}{2010}).

\bibitem{Gauduel:2010}
\bibinfo{author}{Gauduel, YA}, \bibinfo{author}{Glinec, Y},
  \bibinfo{author}{Rousseau, JP}, \bibinfo{author}{Burgy, F} \&
  \bibinfo{author}{Malka, V}.
\newblock \bibinfo{title}{{High energy radiation femtochemistry of water
  molecules: early electron-radical pairs processes}}.
\newblock \emph{\bibinfo{journal}{Eur. Phys. J. D}}
  \textbf{\bibinfo{volume}{60}}, \bibinfo{pages}{121--135}
  (\bibinfo{year}{2010}).

\bibitem{Schlenvoigt:2008}
\bibinfo{author}{Schlenvoigt, HP} \emph{et~al.}
\newblock \bibinfo{title}{A compact synchrotron radiation source driven by a
  laser-plasma wakefield accelerator}.
\newblock \emph{\bibinfo{journal}{Nat. Phys.}} \textbf{\bibinfo{volume}{4}},
  \bibinfo{pages}{130--133} (\bibinfo{year}{2008}).

\bibitem{Fuchs:2009}
\bibinfo{author}{Fuchs, M} \emph{et~al.}
\newblock \bibinfo{title}{Laser-driven soft-x-ray undulator source}.
\newblock \emph{\bibinfo{journal}{Nat. Phys.}} \textbf{\bibinfo{volume}{5}},
  \bibinfo{pages}{826--829} (\bibinfo{year}{2009}).

\bibitem{Kneip:2010}
\bibinfo{author}{Kneip, S} \emph{et~al.}
\newblock \bibinfo{title}{{Bright spatially coherent synchrotron X-rays from a
  table-top source}}.
\newblock \emph{\bibinfo{journal}{Nat. Phys.}} \textbf{\bibinfo{volume}{6}},
  \bibinfo{pages}{980--983} (\bibinfo{year}{2010}).

\bibitem{Cipiccia:2011}
\bibinfo{author}{Cipiccia, S} \emph{et~al.}
\newblock \bibinfo{title}{{Gamma-rays from harmonically resonant betatron
  oscillations in a plasma wake}}.
\newblock \emph{\bibinfo{journal}{Nat. Phys.}} \textbf{\bibinfo{volume}{7}},
  \bibinfo{pages}{1--5} (\bibinfo{year}{2011}).

\bibitem{Kneip:2011}
\bibinfo{author}{Kneip, S} \emph{et~al.}
\newblock \bibinfo{title}{X-ray phase contrast imaging of biological specimens
  with femtosecond pulses of betatron radiation from a compact laser plasma
  wakefield accelerator}.
\newblock \emph{\bibinfo{journal}{Appl. Phys. Lett.}}
  \textbf{\bibinfo{volume}{99}}, \bibinfo{pages}{093701}
  (\bibinfo{year}{2011}).

\bibitem{Fourmaux:2011}
\bibinfo{author}{Fourmaux, S} \emph{et~al.}
\newblock \bibinfo{title}{{Single shot phase contrast imaging using
  laser-produced betatron x-ray beams}}.
\newblock \emph{\bibinfo{journal}{Opt. Lett.}} \textbf{\bibinfo{volume}{36}},
  \bibinfo{pages}{2426--2428} (\bibinfo{year}{2011}).

\bibitem{Emma:2010}
\bibinfo{author}{Emma, P} \emph{et~al.}
\newblock \bibinfo{title}{{First lasing and operation of an Angstrom-wavelength
  free-electron laser}}.
\newblock \emph{\bibinfo{journal}{Nat. Photonics}}
  \textbf{\bibinfo{volume}{4}}, \bibinfo{pages}{641--647}
  (\bibinfo{year}{2010}).

\bibitem{Gruner:2007}
\bibinfo{author}{Gr{\"u}ner, FJ} \emph{et~al.}
\newblock \bibinfo{title}{Design considerations for table-top, laser-based vuv
  and x-ray free electron lasers}.
\newblock \emph{\bibinfo{journal}{Appl. Phys. B}}
  \textbf{\bibinfo{volume}{86}}, \bibinfo{pages}{431--435}
  (\bibinfo{year}{2007}).

\bibitem{Huang:2012}
\bibinfo{author}{Huang, Z}, \bibinfo{author}{Ding, Y} \&
  \bibinfo{author}{Schroeder, CB}.
\newblock \bibinfo{title}{{Compact X-ray Free-Electron Laser from a
  Laser-Plasma Accelerator Using a Transverse-Gradient Undulator}}.
\newblock \emph{\bibinfo{journal}{Phys. Rev. Lett.}}
  \textbf{\bibinfo{volume}{109}}, \bibinfo{pages}{204801}
  (\bibinfo{year}{2012}).

\bibitem{Maier:2012}
\bibinfo{author}{Maier, A} \emph{et~al.}
\newblock \bibinfo{title}{{Demonstration Scheme for a Laser-Plasma-Driven
  Free-Electron Laser}}.
\newblock \emph{\bibinfo{journal}{Phys. Rev. X}} \textbf{\bibinfo{volume}{2}},
  \bibinfo{pages}{031019} (\bibinfo{year}{2012}).

\bibitem{Nakajima:2011}
\bibinfo{author}{Nakajima, K} \emph{et~al.}
\newblock \bibinfo{title}{{Operating plasma density issues on large-scale
  laser-plasma accelerators toward high-energy frontier}}.
\newblock \emph{\bibinfo{journal}{Phys. Rev. Spec. Top.--Accel. Beams}}
  \textbf{\bibinfo{volume}{14}} (\bibinfo{year}{2011}).

\bibitem{Schroeder:2012}
\bibinfo{author}{Schroeder, C}, \bibinfo{author}{Esarey, E} \&
  \bibinfo{author}{Leemans, W}.
\newblock \bibinfo{title}{{Beamstrahlung considerations in
  laser-plasma-accelerator-based linear colliders}}.
\newblock \emph{\bibinfo{journal}{Phys. Rev. Spec. Top.--Accel. Beams}}
  \textbf{\bibinfo{volume}{15}}, \bibinfo{pages}{051301}
  (\bibinfo{year}{2012}).

\bibitem{Tilborg:2006}
\bibinfo{author}{van Tilborg, J} \emph{et~al.}
\newblock \bibinfo{title}{Temporal characterization of femtosecond
  laser-plasma-accelerated electron bunches using terahertz radiation}.
\newblock \emph{\bibinfo{journal}{Phys. Rev. Lett.}}
  \textbf{\bibinfo{volume}{96}}, \bibinfo{pages}{014801}
  (\bibinfo{year}{2006}).

\bibitem{Ohkubo:2007}
\bibinfo{author}{Ohkubo, T} \emph{et~al.}
\newblock \bibinfo{title}{{Temporal characteristics of monoenergetic electron
  beams generated by the laser wakefield acceleration}}.
\newblock \emph{\bibinfo{journal}{Phys. Rev. Spec. Top.--Accel. Beams}}
  \textbf{\bibinfo{volume}{10}}, \bibinfo{pages}{031301}
  (\bibinfo{year}{2007}).

\bibitem{Debus:2010}
\bibinfo{author}{Debus, AD} \emph{et~al.}
\newblock \bibinfo{title}{{Electron Bunch Length Measurements from
  Laser-Accelerated Electrons Using Single-Shot THz Time-Domain
  Interferometry}}.
\newblock \emph{\bibinfo{journal}{Phys. Rev. Lett.}}
  \textbf{\bibinfo{volume}{104}}, \bibinfo{pages}{084802}
  (\bibinfo{year}{2010}).

\bibitem{Lundh:2011}
\bibinfo{author}{Lundh, O} \emph{et~al.}
\newblock \bibinfo{title}{{Few femtosecond, few kiloampere electron bunch
  produced by a laser-plasma accelerator}}.
\newblock \emph{\bibinfo{journal}{Nat. Phys.}} \textbf{\bibinfo{volume}{7}},
  \bibinfo{pages}{219--222} (\bibinfo{year}{2011}).

\bibitem{Fritzler:2004}
\bibinfo{author}{Fritzler, S} \emph{et~al.}
\newblock \bibinfo{title}{{Emittance measurements of a
  laser-wakefield-accelerated electron beam}}.
\newblock \emph{\bibinfo{journal}{Phys. Rev. Lett.}}
  \textbf{\bibinfo{volume}{92}}, \bibinfo{pages}{165006}
  (\bibinfo{year}{2004}).

\bibitem{Sears:2010}
\bibinfo{author}{Sears, CMS} \emph{et~al.}
\newblock \bibinfo{title}{{Emittance and divergence of laser wakefield
  accelerated electrons}}.
\newblock \emph{\bibinfo{journal}{Phys. Rev. Spec. Top.--Accel. Beams}}
  \textbf{\bibinfo{volume}{13}}, \bibinfo{pages}{092803}
  (\bibinfo{year}{2010}).

\bibitem{Brunetti:2010}
\bibinfo{author}{Brunetti, E} \emph{et~al.}
\newblock \bibinfo{title}{{Low Emittance, High Brilliance Relativistic Electron
  Beams from a Laser-Plasma Accelerator}}.
\newblock \emph{\bibinfo{journal}{Phys. Rev. Lett.}}
  \textbf{\bibinfo{volume}{{105}}} (\bibinfo{year}{{2010}}).

\bibitem{Plateau:2012}
\bibinfo{author}{Plateau, G} \emph{et~al.}
\newblock \bibinfo{title}{{Low-Emittance Electron Bunches from a Laser-Plasma
  Accelerator Measured using Single-Shot X-Ray Spectroscopy}}.
\newblock \emph{\bibinfo{journal}{Phys. Rev. Lett.}}
  \textbf{\bibinfo{volume}{109}}, \bibinfo{pages}{064802}
  (\bibinfo{year}{2012}).

\bibitem{Weingartner:2012}
\bibinfo{author}{Weingartner, R} \emph{et~al.}
\newblock \bibinfo{title}{{Ultralow emittance electron beams from a
  laser-wakefield accelerator}}.
\newblock \emph{\bibinfo{journal}{Phys. Rev. Spec. Top.--Accel. Beams}}
  \textbf{\bibinfo{volume}{15}}, \bibinfo{pages}{111302}
  (\bibinfo{year}{2012}).

\bibitem{Kneip:2012}
\bibinfo{author}{Kneip, S} \emph{et~al.}
\newblock \bibinfo{title}{{Characterization of transverse beam emittance of
  electrons from a laser-plasma wakefield accelerator in the bubble regime
  using betatron x-ray radiation}}.
\newblock \emph{\bibinfo{journal}{Phys. Rev. Spec. Top.--Accel. Beams}}
  \textbf{\bibinfo{volume}{15}}, \bibinfo{pages}{021302}
  (\bibinfo{year}{2012}).

\bibitem{Esarey:1997}
\bibinfo{author}{Esarey, E}, \bibinfo{author}{Hubbard, RF},
  \bibinfo{author}{Leemans, WP}, \bibinfo{author}{Ting, A} \&
  \bibinfo{author}{Sprangle, P}.
\newblock \bibinfo{title}{Electron injection into plasma wake fields by
  colliding laser pulses}.
\newblock \emph{\bibinfo{journal}{Phys. Rev. Lett.}}
  \textbf{\bibinfo{volume}{79}}, \bibinfo{pages}{2682--2685}
  (\bibinfo{year}{1997}).

\bibitem{Faure:2006}
\bibinfo{author}{Faure, J} \emph{et~al.}
\newblock \bibinfo{title}{Controlled injection and acceleration of electrons in
  plasma wakefields by colliding laser pulses}.
\newblock \emph{\bibinfo{journal}{Nature}} \textbf{\bibinfo{volume}{444}},
  \bibinfo{pages}{737--739} (\bibinfo{year}{2006}).

\bibitem{Bulanov:1998}
\bibinfo{author}{Bulanov, S}, \bibinfo{author}{Naumova, N},
  \bibinfo{author}{Pegoraro, F} \& \bibinfo{author}{Sakai, J}.
\newblock \bibinfo{title}{Particle injection into the wave acceleration phase
  due to nonlinear wake wave breaking}.
\newblock \emph{\bibinfo{journal}{Phys. Rev. E}} \textbf{\bibinfo{volume}{58}},
  \bibinfo{pages}{R5257--R5260} (\bibinfo{year}{1998}).

\bibitem{Brantov:2008}
\bibinfo{author}{Brantov, AV} \emph{et~al.}
\newblock \bibinfo{title}{Controlled electron injection into the wake wave
  using plasma density inhomogeneity}.
\newblock \emph{\bibinfo{journal}{Phys. Plasmas}}
  \textbf{\bibinfo{volume}{15}}, \bibinfo{pages}{073111}
  (\bibinfo{year}{2008}).

\bibitem{Geddes:2008}
\bibinfo{author}{Geddes, CGR} \emph{et~al.}
\newblock \bibinfo{title}{Plasma-density-gradient injection of low
  absolute-momentum-spread electron bunches}.
\newblock \emph{\bibinfo{journal}{Phys. Rev. Lett.}}
  \textbf{\bibinfo{volume}{100}}, \bibinfo{pages}{215004}
  (\bibinfo{year}{2008}).

\bibitem{Faure:2010}
\bibinfo{author}{Faure, J}, \bibinfo{author}{Rechatin, C},
  \bibinfo{author}{Lundh, O}, \bibinfo{author}{Ammoura, L} \&
  \bibinfo{author}{Malka, V}.
\newblock \bibinfo{title}{{Injection and acceleration of quasimonoenergetic
  relativistic electron beams using density gradients at the edges of a plasma
  channel}}.
\newblock \emph{\bibinfo{journal}{Phys. Plasmas}}
  \textbf{\bibinfo{volume}{17}}, \bibinfo{pages}{083107}
  (\bibinfo{year}{2010}).

\bibitem{Suk:2001}
\bibinfo{author}{Suk, H}, \bibinfo{author}{Barov, N},
  \bibinfo{author}{Rosenzweig, JB} \& \bibinfo{author}{Esarey, E}.
\newblock \bibinfo{title}{Plasma electron trapping and acceleration in a plasma
  wake field using a density transition}.
\newblock \emph{\bibinfo{journal}{Phys. Rev. Lett.}}
  \textbf{\bibinfo{volume}{86}}, \bibinfo{pages}{1011--1014}
  (\bibinfo{year}{2001}).

\bibitem{Schmid:2010}
\bibinfo{author}{Schmid, K} \emph{et~al.}
\newblock \bibinfo{title}{{Density-transition based electron injector for laser
  driven wakefield accelerators}}.
\newblock \emph{\bibinfo{journal}{Phys. Rev. Spec. Top.--Accel. Beams}}
  \textbf{\bibinfo{volume}{13}}, \bibinfo{pages}{091301}
  (\bibinfo{year}{2010}).

\bibitem{Rowlands-Rees:2008}
\bibinfo{author}{Rowlands-Rees, TP} \emph{et~al.}
\newblock \bibinfo{title}{Laser-driven acceleration of electrons in a partially
  ionized plasma channel}.
\newblock \emph{\bibinfo{journal}{Phys. Rev. Lett.}}
  \textbf{\bibinfo{volume}{100}}, \bibinfo{pages}{105005}
  (\bibinfo{year}{2008}).

\bibitem{Pak:2010}
\bibinfo{author}{Pak, A} \emph{et~al.}
\newblock \bibinfo{title}{Injection and trapping of tunnel-ionized electrons
  into laser-produced wakes}.
\newblock \emph{\bibinfo{journal}{Phys. Rev. Lett.}}
  \textbf{\bibinfo{volume}{104}}, \bibinfo{pages}{025003}
  (\bibinfo{year}{2010}).

\bibitem{McGuffey:2010}
\bibinfo{author}{McGuffey, C} \emph{et~al.}
\newblock \bibinfo{title}{Ionization induced trapping in a laser wakefield
  accelerator}.
\newblock \emph{\bibinfo{journal}{Phys. Rev. Lett.}}
  \textbf{\bibinfo{volume}{104}}, \bibinfo{pages}{025004}
  (\bibinfo{year}{2010}).

\bibitem{Liu:2011}
\bibinfo{author}{Liu, J} \emph{et~al.}
\newblock \bibinfo{title}{{All-Optical Cascaded Laser Wakefield Accelerator
  Using Ionization-Induced Injection}}.
\newblock \emph{\bibinfo{journal}{Phys. Rev. Lett.}}
  \textbf{\bibinfo{volume}{107}}, \bibinfo{pages}{035001}
  (\bibinfo{year}{2011}).

\bibitem{Panasenko:2010}
\bibinfo{author}{Panasenko, D} \emph{et~al.}
\newblock \bibinfo{title}{Demonstration of a plasma mirror based on a laminar
  flow water film}.
\newblock \emph{\bibinfo{journal}{J. Appl. Phys.}}
  \textbf{\bibinfo{volume}{108}}, \bibinfo{pages}{044913}
  (\bibinfo{year}{2010}).

\bibitem{Sprangle:2001}
\bibinfo{author}{Sprangle, P} \emph{et~al.}
\newblock \bibinfo{title}{Wakefield generation and gev acceleration in tapered
  plasma channels}.
\newblock \emph{\bibinfo{journal}{Phys. Rev. E}} \textbf{\bibinfo{volume}{63}},
  \bibinfo{pages}{056405} (\bibinfo{year}{2001}).

\bibitem{Pukhov:2008}
\bibinfo{author}{Pukhov, A} \& \bibinfo{author}{Kostyukov, I}.
\newblock \bibinfo{title}{Control of laser-wakefield acceleration by the
  plasma-density profile}.
\newblock \emph{\bibinfo{journal}{Phys. Rev. E}} \textbf{\bibinfo{volume}{77}},
  \bibinfo{pages}{025401} (\bibinfo{year}{2008}).

\bibitem{ICFA56}
\bibinfo{author}{Leemans, W}, \bibinfo{author}{Chou, W} \&
  \bibinfo{author}{Uesaka, M}.
\newblock \bibinfo{title}{White paper of the icfa-icuil joint task force {---}
  high power laser technology for accelerators}.
\newblock \emph{\bibinfo{journal}{ICFA Beam Dynamics Newsletter}}
  \textbf{\bibinfo{volume}{56}}, \bibinfo{pages}{10--87}
  (\bibinfo{year}{2011}).

\bibitem{Dawson:2012}
\bibinfo{author}{Dawson, JW} \emph{et~al.}
\newblock \bibinfo{title}{High average power lasers for future particle
  accelerators}.
\newblock \emph{\bibinfo{journal}{AIP Conf. Proc.}}
  \textbf{\bibinfo{volume}{1507}}, \bibinfo{pages}{147--153}
  (\bibinfo{year}{2012}).

\bibitem{Richardson:2010}
\bibinfo{author}{Richardson, DJ}, \bibinfo{author}{Nilsson, J} \&
  \bibinfo{author}{Clarkson, WA}.
\newblock \bibinfo{title}{High power fiber lasers: current status and future
  perspectives}.
\newblock \emph{\bibinfo{journal}{J. Opt. Soc. Am. B}}
  \textbf{\bibinfo{volume}{27}}, \bibinfo{pages}{B63--B92}
  (\bibinfo{year}{2010}).

\bibitem{Ross:1997}
\bibinfo{author}{Ross, I}, \bibinfo{author}{Matousek, P},
  \bibinfo{author}{Towrie, M}, \bibinfo{author}{Langley, A} \&
  \bibinfo{author}{Collier, J}.
\newblock \bibinfo{title}{The prospects for ultrashort pulse duration and
  ultrahigh intensity using optical parametric chirped pulse amplifiers}.
\newblock \emph{\bibinfo{journal}{Opt. Commun.}}
  \textbf{\bibinfo{volume}{144}}, \bibinfo{pages}{125 -- 133}
  (\bibinfo{year}{1997}).

\bibitem{Dubietis:2006}
\bibinfo{author}{Dubietis, A}, \bibinfo{author}{Butkus, R} \&
  \bibinfo{author}{Piskarskas, AP}.
\newblock \bibinfo{title}{{Trends in chirped pulse optical parametric
  amplification}}.
\newblock \emph{\bibinfo{journal}{IEEE J. Sel. Top. Quantum Electron.}}
  \textbf{\bibinfo{volume}{12}}, \bibinfo{pages}{163--172}
  (\bibinfo{year}{2006}).

\bibitem{Goodno:2006}
\bibinfo{author}{Goodno, GD} \emph{et~al.}
\newblock \bibinfo{title}{Coherent combination of high-power, zigzag slab
  lasers}.
\newblock \emph{\bibinfo{journal}{Opt. Lett.}} \textbf{\bibinfo{volume}{31}},
  \bibinfo{pages}{1247--1249} (\bibinfo{year}{2006}).

\bibitem{Krauss:2009}
\bibinfo{author}{Krauss, G} \emph{et~al.}
\newblock \bibinfo{title}{{Synthesis of a single cycle of light with compact
  erbium-doped fibre technology}}.
\newblock \emph{\bibinfo{journal}{Nat. Photonics}} \bibinfo{pages}{1--4}
  (\bibinfo{year}{2009}).

\bibitem{Eidam:2011}
\bibinfo{author}{Eidam, T} \emph{et~al.}
\newblock \bibinfo{title}{Fiber chirped-pulse amplification system emitting
  {3.8\,GW} peak power}.
\newblock \emph{\bibinfo{journal}{Opt. Express}} \textbf{\bibinfo{volume}{19}},
  \bibinfo{pages}{255--260} (\bibinfo{year}{2011}).

\bibitem{Morou:2013}
\bibinfo{author}{Mourou, G}, \bibinfo{author}{Brocklesby, B},
  \bibinfo{author}{Tajima, T} \& \bibinfo{author}{Limpert, J}.
\newblock \bibinfo{title}{The future is fibre accelerators}.
\newblock \emph{\bibinfo{journal}{Nat. Photonics}}
  \textbf{\bibinfo{volume}{7}}, \bibinfo{pages}{258--261}
  (\bibinfo{year}{2013}).

\bibitem{Tamovsauskas:2008}
\bibinfo{author}{Tamosauskas, G}, \bibinfo{author}{Dubietis, A},
  \bibinfo{author}{Valiulis, G} \& \bibinfo{author}{Piskarskas, A}.
\newblock \bibinfo{title}{{Optical parametric amplifier pumped by two mutually
  incoherent laser beams}}.
\newblock \emph{\bibinfo{journal}{Appl. Phys. B}}
  \textbf{\bibinfo{volume}{91}}, \bibinfo{pages}{305--307}
  (\bibinfo{year}{2008}).

\bibitem{Alisauskas:2010}
\bibinfo{author}{Alisauskas, S} \emph{et~al.}
\newblock \bibinfo{title}{{Prospects for increasing average power of optical
  parametric chirped pulse amplifiers via multi-beam pumping}}.
\newblock \emph{\bibinfo{journal}{Opt. Commun.}}
  \textbf{\bibinfo{volume}{283}}, \bibinfo{pages}{469--473}
  (\bibinfo{year}{2010}).

\bibitem{Kurita:2010}
\bibinfo{author}{Kurita, T}, \bibinfo{author}{Sueda, K},
  \bibinfo{author}{Tsubakimoto, K} \& \bibinfo{author}{Miyanaga, N}.
\newblock \bibinfo{title}{Experimental demonstration of spatially coherent beam
  combining using optical parametric amplification}.
\newblock \emph{\bibinfo{journal}{Opt. Express}} \textbf{\bibinfo{volume}{18}},
  \bibinfo{pages}{14541--14546} (\bibinfo{year}{2010}).

\bibitem{Herrmann:2009}
\bibinfo{author}{Herrmann, D} \emph{et~al.}
\newblock \bibinfo{title}{{Generation of sub-three-cycle, 16 TW light pulses by
  using noncollinear optical parametric chirped-pulse amplification}}.
\newblock \emph{\bibinfo{journal}{Opt. Lett.}} \textbf{\bibinfo{volume}{34}},
  \bibinfo{pages}{2459--2461} (\bibinfo{year}{2009}).

\bibitem{Lozhkarev:2007}
\bibinfo{author}{Lozhkarev, VV} \emph{et~al.}
\newblock \bibinfo{title}{{Compact 0.56 Petawatt laser system based on optical
  parametric chirped pulse amplification in KD*P crystals}}.
\newblock \emph{\bibinfo{journal}{Laser Phys. Lett.}}
  \textbf{\bibinfo{volume}{4}}, \bibinfo{pages}{421--427}
  (\bibinfo{year}{2007}).

\bibitem{Skrobol:2012}
\bibinfo{author}{Skrobol, C} \emph{et~al.}
\newblock \bibinfo{title}{{Broadband amplification by picosecond OPCPA in DKDP
  pumped at 515 nm}}.
\newblock \emph{\bibinfo{journal}{Optics Express}}
  \textbf{\bibinfo{volume}{20}}, \bibinfo{pages}{4619--4629}
  (\bibinfo{year}{2012}).

\bibitem{Nakajima:1992}
\bibinfo{author}{Nakajima, K}.
\newblock \bibinfo{title}{Plasma-wave resonator for particle-beam
  acceleration}.
\newblock \emph{\bibinfo{journal}{Phys. Rev. A}} \textbf{\bibinfo{volume}{45}},
  \bibinfo{pages}{1149--1156} (\bibinfo{year}{1992}).

\bibitem{Umstadter:1994}
\bibinfo{author}{Umstadter, D}, \bibinfo{author}{Esarey, E} \&
  \bibinfo{author}{Kim, J}.
\newblock \bibinfo{title}{Nonlinear plasma waves resonantly driven by optimized
  laser pulse trains}.
\newblock \emph{\bibinfo{journal}{Phys. Rev. Lett.}}
  \textbf{\bibinfo{volume}{72}}, \bibinfo{pages}{1224--1227}
  (\bibinfo{year}{1994}).

\bibitem{Corner:2012}
\bibinfo{author}{Corner, L} \emph{et~al.}
\newblock \bibinfo{title}{Multiple pulse resonantly enhanced laser plasma
  wakefield acceleration}.
\newblock \emph{\bibinfo{journal}{AIP Conf. Proc.}}
  \textbf{\bibinfo{volume}{1507}}, \bibinfo{pages}{872--873}
  (\bibinfo{year}{2012}).

\end{thebibliography}
\end{document}